\documentclass[preprint,12pt]{emulateapj}
\usepackage{xspace}
\usepackage{mathrsfs}

\newcommand{\kep}{{\it Kepler}\xspace}
\newcommand{\Kepler}{{\it Kepler}}
\newcommand{\tess}{{\it TESS}}

\newcommand{\epic}{EPIC 246851721\xspace}
\newcommand{\ktwo}{{\it K2}\xspace}

\newcommand{\be}{\begin{equation}}
\newcommand{\ee}{\end{equation}}

\newcommand{\msun}{\ensuremath{M_\odot}}
\newcommand{\rsun}{\ensuremath{R_\odot}}
\newcommand{\mj}{M$_\mathrm{J}$\xspace}
\newcommand{\rj}{R$_\mathrm{J}$\xspace}

\newcommand{\kms}{\ensuremath{\rm km\,s^{-1}}}
\newcommand{\ms}{\ensuremath{\rm m\,s^{-1}}}

\newcommand{\bjdtdb}{\ensuremath{\rm {BJD_{TDB}}}}
\newcommand{\feh}{\ensuremath{\left[{\rm Fe}/{\rm H}\right]}}

\newcommand{\lsun}{\ensuremath{\,L_\Sun}}

\newcommand{\fave}{\langle F \rangle}
\newcommand{\fluxcgs}{10$^9$ erg s$^{-1}$ cm$^{-2}$}
\newcommand{\teff}{$T_\mathrm{eff}$\xspace}

\newcommand{\logg}{$\log g$\xspace}

\newcommand{\umpl}{1.1}
\newcommand{\mpl}{3.0}
\newcommand{\dmpl}{1.2}

\newcommand{\rpl}{1.051\xspace}
\newcommand{\rple}{0.044}

\newcommand{\vsini}{\ensuremath{v \sin I_\star}\xspace}
\newcommand{\vsinival}{74.92}

\newcommand{\mstar}{1.317}

\newcommand{\rstar}{1.586}

\newcommand{\planet}{EPIC~246851721~b\xspace}

\usepackage{xcolor}
\usepackage{hyperref}
\usepackage{cleveref}
\usepackage{breakurl}
\usepackage{amsmath}
\usepackage{graphicx}
\usepackage{verbatim}
\usepackage{booktabs}
\usepackage{longtable}

\slugcomment{}

\shorttitle{Doppler Tomography of a Tropical Jupiter}
\shortauthors{Yu et al.}

\begin{document}


\title{EPIC 246851721 \MakeLowercase{b}: A Tropical Jupiter Transiting a Rapidly Rotating Star in a Well-Aligned Orbit}
\author{Liang Yu\altaffilmark{1}, George Zhou\altaffilmark{2,$\star$}, Joseph E. Rodriguez\altaffilmark{2}, Chelsea X. Huang\altaffilmark{1},  Andrew Vanderburg\altaffilmark{3,$\dagger$}, Samuel N. Quinn\altaffilmark{2}, B. Scott Gaudi\altaffilmark{4}, Charles A. Beichman\altaffilmark{5}, Perry Berlind\altaffilmark{2}, Allyson Bieryla\altaffilmark{2},  Michael L. Calkins\altaffilmark{2}, David R. Ciardi\altaffilmark{5}, Ian J. M. Crossfield\altaffilmark{1}, Jason D. Eastman\altaffilmark{2}, Gilbert A. Esquerdo\altaffilmark{2}, David W. Latham\altaffilmark{2},  Keivan G. Stassun\altaffilmark{6}, Steven Villanueva Jr.\altaffilmark{1,4}}

\altaffiltext{1}{ Department of Physics, and Kavli Institute for Astrophysics and Space Research, Massachusetts Institute of Technology, Cambridge, MA 02139, USA}
\altaffiltext{2}{Harvard--Smithsonian Center for Astrophysics, 60 Garden St., Cambridge, MA 02138, USA}
\altaffiltext{3}{Department of Astronomy, The University of Texas at Austin, Austin, TX 78712, USA}
\altaffiltext{4}{Department of Astronomy, The Ohio State University, Columbus, OH 43210, USA}
\altaffiltext{5}{NASA Exoplanet Science Institute, California Institute of Technology, Jet Propulsion Laboratory, Pasadena, CA 91125, USA}
\altaffiltext{6}{Vanderbilt University, Department of Physics \& Astronomy, 6301 Stevenson Center Ln., Nashville, TN 37235, USA}
\altaffiltext{$\star$}{NASA Hubble Fellow}
\altaffiltext{$\dagger$}{NASA Sagan Fellow}


\begin{abstract}
We report the discovery of \planet, a ``tropical" Jupiter in a 6.18-day orbit around the bright ($V=11.439$) star \epic (TYC 1283-739-1). We present a detailed analysis of the system using \ktwo and ground-based photometry, radial velocities, Doppler tomography and adaptive optics imaging. From our global models, we infer that the host star is a rapidly rotating (\vsini = \vsinival \ \kms) F dwarf with \teff = 6202 K, $R_\star = \rstar \ \rsun$ and $M_\star=\mstar \ \msun$.
\planet has a radius of \rpl$\pm$ \rple \rj, and a mass of \mpl$^{+\umpl}_{-\dmpl}$ \mj. Doppler tomography reveals an aligned spin-orbit geometry, with a projected obliquity of $-1.47^{\circ\ +0.87}_{\ -0.86}$, making \epic the fourth hottest star to host a Jovian planet with $P > 5$ days and a known obliquity. Using quasi-periodic signatures in its light curve that appear to be spot modulations, we estimate the star's rotation period, and thereby infer the true obliquity of the system to be $3.7^{\circ\ +3.7}_{\ -1.8}$. We argue that this near-zero obliquity is likely to be primordial rather than a result of tidal damping. The host star also has a bound stellar companion, a 0.4 \msun \ M dwarf at a projected separation of 2100 AU, but the companion is likely incapable of emplacing \planet in its current orbit via high eccentricity Kozai-Lidov migration.

\end{abstract}


\section{Introduction}
Ever since their initial discovery in 1995, the formation and evolution history of hot Jupiters have been the subjects of intense study. We have yet to understand how hot Jupiters, giant planets orbiting at just a few percent of an astronomical unit from their host stars, are able to reach such close-in orbits. Their origin theories fall into three categories: {\it in situ} formation \citep[e.g.][]{bodenheimer00, batygin16,Boley:2016}, high eccentricity migration via gravitational perturbations by bound stellar companions \citep[e.g.][]{wumurray} or by other planets in the system \citep[e.g.][]{rasioford, naoz11}, and smooth disk migration \citep[e.g.][]{lin96, alibert05}. Among these mechanisms, high eccentricity migration is capable of altering the planet's orbital orientation and driving up the stellar obliquity (i.e. the angle between the planetary orbit and the host star's spin axis), while {\it in situ} formation and disk migration are likely to preserve or even reduce the initial obliquity, although this initial obliquity may not necessarily be small \citep[see e.g.][]{Batygin:2013,lai14}. Therefore the measurement of stellar obliquities is a key to understanding the migration of hot Jupiters \citep[e.g.][]{albrecht12}. Only the sky projection of this angle, $\lambda$, can be directly measured via the Rossiter-McLaughlin effect \citep{McLaughlin:1924, Rossiter:1924}, Doppler tomography \citep{Cameron:2010}, or star spot crossings \citep{sanchisojeda11}, among other techniques. As of June 2018, 125 planets have measured projected obliquities\footnote{http://www.astro.keele.ac.uk/jkt/tepcat/obliquity.html}, the majority of which are hot Jupiters with orbital periods shorter than 5 days. The projected obliquity is only a lower bound on the true, three-dimensional obliquity, of which only 23 planets have measurements.

As was first noted by \citet{winn10}, \citet{schlaufman10} and others, among hot Jupiter hosts, stars with cool photospheres ($\lesssim 6100$ K) are generally in spin-orbit alignment, while hotter stars ($\gtrsim 6100$ K) span a wide range of obliquities. The temperature boundary coincides with the ``Kraft break", the transition between convective and radiative stars \citep{kraft67}. This dichotomy is commonly thought to arise from differences in internal structures between hot and cool stars. Under this hypothesis, hot Jupiter systems acquire a wide range of obliquities through the migration process of the planet \citep[e.g.][]{rasioford, fabryckytremaine}, magnetic star-disk interactions \citep{lai14}, or torques from distant stellar companions \citep{batygin12}. But stars cooler than $\sim 6100$ K have thick convective envelopes and high rates of tidal dissipation, and are able to quickly realign orbits through tidal interactions. Hot stars, on the other hand, can only weakly dampen orbital obliquities and cannot realign the planets on Gyr timescales. 

Recently, however, this tidal realignment hypothesis has been challenged by new observations. Using ensemble measurements, \citet{mazeh15} found that the hot/cool obliquity distinction persists even for long-period planets, where tidal interactions should be negligible. Theoretically, it is also difficult to explain how a planet can realign its host star without sacrificing all of its angular momentum and leading to orbital decay \citep[e.g.][]{lai12, damianilanza, linogilvie}. Therefore the interpretation of hot Jupiters' obliquities remains an outstanding problem, which calls for an expansion of the parameter space for which we have obliquity measurements. In particular, we need to measure the obliquities of planets with longer orbital periods than typical hot Jupiters, including warm Jupiters (with orbital periods $\gtrsim 10$ days) and Jovian planets with periods between 5 and 10 days, which we dub ``tropical" Jupiters. 

Unlike typical hot Jupiters, which have periods shorter than 5 days, tropical Jupiters may provide important insight into the dynamic evolution of close-in giant planets. At farther distances from their host stars, tropical Jupiters experience weaker tidal effects, and thus may have retained both the obliquity and eccentricity they had when emplaced in their current orbits. The tidal circularization timescale scales with the orbital period as $\tau_\mathrm{circ} \propto P^{13/3}$ \citep{adamslaughlin}. For a tropical Jupiter with a period of 5 days, the circularization timescale is as long as $\sim 2$ Gyr. This means that if close-in Jovian planets arrive through high eccentricity migration, we should be able to detect non-zero primordial eccentricities and obliquities in tropical Jupiters. Tropical Jupiters are also distinct from warm Jupiters, which are often defined in the literature as giant planets with incident irradiation levels below $2\times10^8$ erg s$^{-1}$ cm$^{-2}$.

In this paper, we present the discovery and obliquity measurement of a tropical Jupiter, \planet, using Doppler tomography. The planet transits a rapidly rotating (\vsini = \vsinival \ \kms), bright F dwarf on a 6.18-day orbit. Despite having an effective temperature just above the Kraft break, the planet appears to be very well aligned. Using what appears to be spot modulations in the \ktwo light curve, we were able to estimate the true obliquity of this system. Interestingly, the host star also has a distant, bound stellar companion, which enables us to examine the role of stellar Kozai-Lidov oscillations in the planet's migration history. This paper is organized as follows: we describe our discovery and observations in Section~\ref{observations}, our derivation of stellar and planetary parameters in Sections~\ref{star} and~\ref{sec:exofast}, and our interpretation of the stellar companion and well-aligned planetary orbit and future prospects in Section~\ref{discussion}.

\section{Observations}\label{observations}
\epic was observed by \kep between March 8 and May 27, 2017 during Campaign 13 of its \ktwo mission. It was proposed as a target in three programs: GO13071 (PI Charbonneau), GO13122 (PI Howard) and GO13024 (PI Cochran). A summary of \epic's photometric and kinematic properties is given in Table~\ref{tab:stellar}.

\subsection{K2 Light Curve}\label{lightcurve}
We extracted photometry for \epic from pixel-level data, which we downloaded from the Mikulski Archive for Space Telescopes (MAST)\footnote{https://archive.stsci.edu/k2/}. We computed the raw aperture photometry by summing the flux inside an irregularly shaped aperture consisting of all pixels that are within some distance away from the centroid of the star. For \epic, a distance of 5 pixels yielded the best-calibrated photometry.

\ktwo photometry is dominated by systematics resulting from spacecraft motion. During spacecraft rolls, stars move over different pixels with varying sensitivities, leading to apparent changes in stellar brightness. As shown by \citet{vj14}, these systematics are strongly correlated with centroid positions of the stars. Our light curve detrending process is similar to that described in detail in \citet{huang15}. In brief, we used the averaged centroid motions of 4 bright stars to represent the centroid positions of all stars in the same campaign. Short-term variations in flux due to spacecraft motion are fitted as the sum of a set of cosine and sine functions of centroid position and the first 10 principal components of fluxes of stars in the same channel. The systematics model is then removed from the raw light curve.  

We then filtered out any remaining low-frequency variability (mostly intrinsic stellar variability) using a set of cosine and sine functions, following the method described in \citet{huang13}. In order to avoid distorting the transit profiles, we set the minimum timescale of the harmonic filter to 0.5 days. We used the box-least-squares (BLS) algorithm \citep{bls} to search for periodic transit signals in the filtered light curve, following procedures outlined in \citet{huang13}, and selected candidates according to statistics produced with the VARTOOLS \citep{vartools} implementation of BLS. This search yielded a signal with a SNR of 61.9 and a preliminary period of 6.18 days.

\begin{figure*}[ht] 
   \centering
   \includegraphics[width=0.8\textwidth]{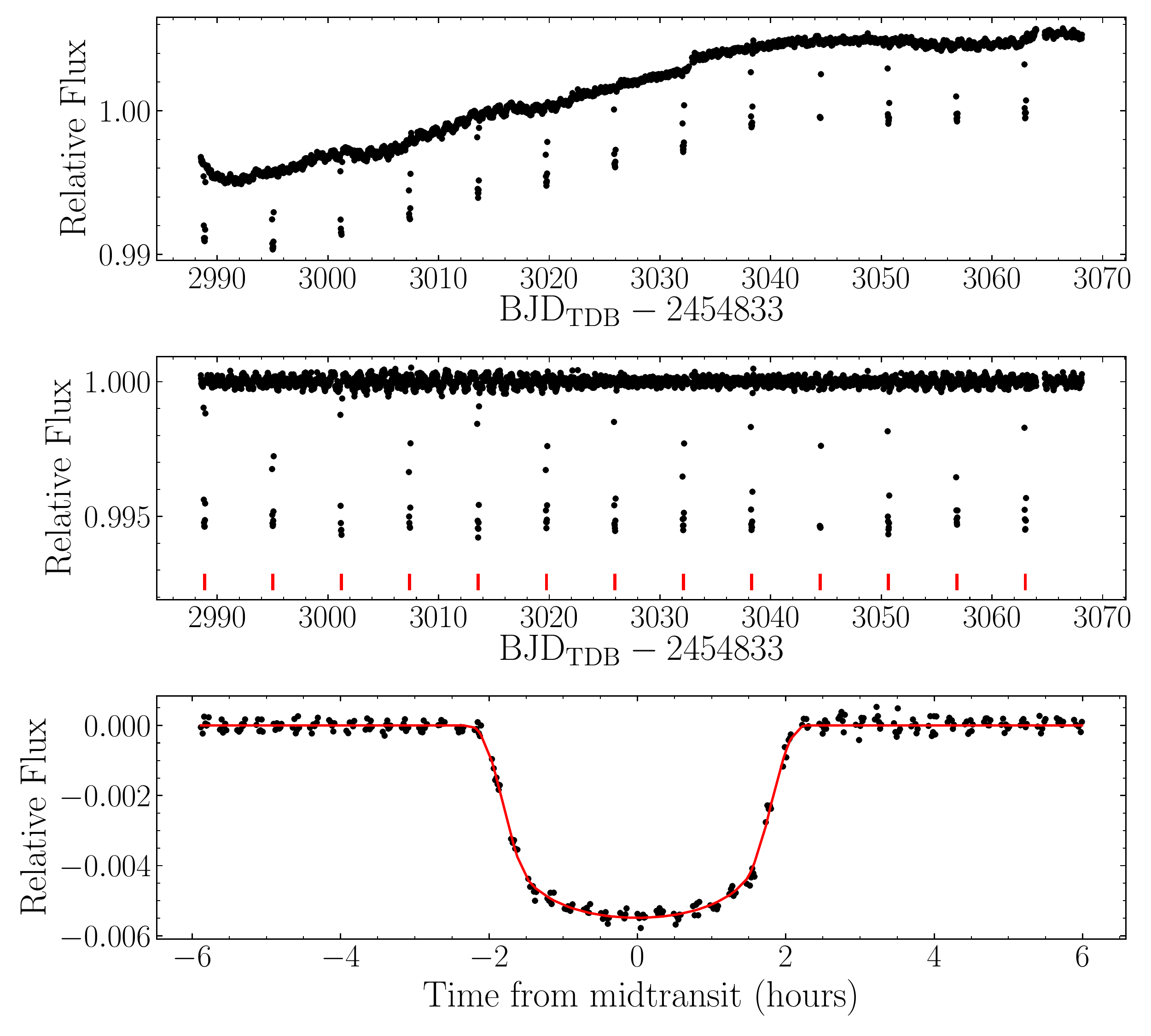} 
   \caption{\ktwo light curves of \epic. From top to bottom: \ktwo photometry after decorrelation against centroid motion; the full light curve after removing low-frequency stellar variability, with vertical ticks indicating the locations of transits (note the residual high-frequency stellar activity in the baseline); the phase-folded photometry and best-fit transit model.}
   \label{fig:lc}
\end{figure*}

After identifying the transits, we produced new light curves by simultaneously fitting the transits, the \ktwo roll systematics, and long-timescale stellar/instrumental variability. Reprocessing the \ktwo light curves in this way prevents the shape of the transits from being biased by the removal of \ktwo systematics. We used light curves and corrections of systematics derived using the method of \citet{vj14} as initial guesses for our simultaneous fits, which we then performed following \citet{v16}. Throughout the rest of this paper, we use these simultaneously-fit light curves in our analysis and our plots. Fig.~\ref{fig:lc} shows the systematics-corrected, flattened\footnote{We flattened the light curve by dividing away the best-fit long-timescale variability from our simultaneously-fit light curve.} and detrended light curves of \epic.

\begin{deluxetable*}{l l l}[bt]
\hspace{-1in}\tabletypesize{\scriptsize}
\tablecaption{  Parameters of \epic from the literature and spectroscopy \label{tab:stellar}}
\tablewidth{0pt}
\tablehead{
\colhead{Parameter}  & \colhead{Value} & \colhead{Source}
}
\startdata
\multicolumn{3}{l}{\hspace{1cm}\em Identifying Information} \\
$\alpha_{\rm J2000}$ R.A. (hh:mm:ss)  & 05:15:40.74 & \\
$\delta_{\rm J2000}$ Dec. (dd:mm:ss) & +16:16:43.47 &  \\
Other identifiers
& TYC 1283-739-1  & \\
& 2MASS J05154075+1616435 & \\
  & K2-YYY \\
\multicolumn{3}{l}{\hspace{1cm}\em Photometric Properties} \\
B (mag)..........  & 11.990 $\pm$ 0.011 & 1\\
V (mag) .......... & 11.439 $\pm$ 0.053 &  1 \\
J (mag)..........  & 10.199 $\pm$ 0.042 &  2 \\
H (mag) .........  & 9.967 $\pm$ 0.021 &  2 \\
K (mag) ........  & 9.893 $\pm$ 0.018 &  2\\
W1 (mag) ........  & 9.813 $\pm$ 0.023 &  3\\
W2 (mag) ........  & 9.816 $\pm$ 0.020 & 3\\
W3 (mag) ........  & 9.754 $\pm$ 0.056 &   3 \\
\multicolumn{3}{l}{\hspace{1cm}\em Spectroscopic and Derived Properties} \\
Spectral Type  & F5V & 4\\
$\mu_{\alpha}$ (mas~yr$^{-1}$) & 2.72 $\pm$ 0.11 & 5 \\
$\mu_{\delta}$ (mas~yr$^{-1}$) & -12.983 $\pm$ 0.094 & 5 \\
Parallax (mas) &  $2.66\pm0.13$ & 5 \\
Barycentric RV (km~s$^{-1}$) &  $36.85^{+0.13}_{-0.12}$  & TRES; this paper \\
\vsini\ (km~s$^{-1}$) & 77.70 $\pm$ 0.78 & TRES; this paper \\
Space motion ($U^{*},V,W$) (km~s$^{-1}$) & ($-22.6 \pm 0.3, -12.6 \pm 1.1, -10.1\pm 0.5$)   & this paper
\enddata
\tablenotetext{}{References: (1) \citet{tycho}; (2) \citet{twomass}; (3) \citet{allwise}; (4) \citet{pecautmamajek13}, (5) \citet{gaia16, gaia18}; $^{*}$Positive $U$ is in the direction of the Galactic center.}
\end{deluxetable*}

\subsection{Ground-Based Follow-Up}
In this section, we present our ground-based photometric and spectroscopic observations used to confirm the planetary nature of \planet. Section~\ref{spectroscopy} describes our radial velocity measurements with TRES and Doppler tomography with MIKE; Section~\ref{sec:ao} presents adaptive optics imaging follow-up with NIRC2; Section~\ref{demonext} describes additional ground-based photometry obtained with the DEMONEXT telescope.

\subsubsection{Spectroscopy}\label{spectroscopy}

To constrain the mass of the planet and measure the stellar atmosphere properties of the host star, we obtained a series of spectroscopic observations with the Tillinghast Reflector Echelle Spectrograph \citep[TRES;][]{tres,Furesz:2008} on the 1.5\,m telescope at Fred Lawrence Whipple Observatory, Mt Hopkins, Arizona. TRES is a fibre-fed echelle spectrograph with a resolving power of $R \equiv \lambda / \Delta \lambda \sim 44,000$, over the wavelength range $3900-9100$\,\AA, spanning 51 echelle orders. A total of 17 observations were obtained with TRES over the time period 2017 September 24 - 2017 November 14, covering multiple orbital phases of the planet. The spectra were reduced as per \citet{buchhave2010}. Relative radial velocities were extracted by cross-correlating against the median observed TRES spectrum. In order to avoid regions of the spectra containing low SNR telluric lines, or little information content, cross-correlations were carried out across only 12 echelle orders in the wavelength range $4290-5485$\,\AA. We track and correct for TRES instrumental zero point shifts using nightly observations of RV standard stars; the shifts ($\lesssim10$ \ms) are small compared to the RV uncertainties of \epic. The final relative velocities are reported in Table~\ref{tab:rv}. To translate the relative RVs to an absolute scale, we calculate two additional offsets: that between the relative and absolute TRES RVs; and that between the absolute TRES and IAU scales. Absolute TRES RVs are calculated by cross-correlating the spectra against synthetic templates, and these absolute TRES RVs are used to calculate the weighted mean offset between relative and absolute TRES RVs. The offset from TRES to the IAU scale is calculated from the absolute TRES velocities of the RV standards. The final absolute center-of-mass velocity for the system is given in Table~\ref{tab:stellar}, but the offsets applied carry additional uncertainties not representative of our sensitivity to orbital motion. For this reason, we perform our analysis on the relative RVs, and report the fitted center-of-mass velocity (i.e., on the relative scale) in Table~\ref{tab:params}. Due to the rapid rotation $(v\sin I_\star = \vsinival\,\mathrm{km\,s}^{-1})$ of the host star, the radial velocities were of lower precision than that typically achieved with TRES. With the velocities, we can only detect the mass of the planet at $2.5\sigma$ significance, with a value of $\mpl^{+\umpl}_{-\dmpl}$ \mj. 

\begin{table}[ht]
  \caption{TRES Radial Velocities\label{tab:rv}}
  \centering
  \begin{tabular}{lrr}
\toprule
        $\mathrm{BJD_{TDB}}$ &     RV  &    $\sigma_{RV}$  \\
 & [m~s$^{-1}$] & [m~s$^{-1}$] \\
        \midrule
 2458020.970114 & -593 & 264\\ 
2458024.001740 & -156 & 440\\ 
2458030.867169 & 182 & 236\\  
2458039.840741 & 167 & 271\\  
2458042.843727 & 1136 & 428\\ 
2458051.906756 & -590 & 287\\ 
2458052.901266 & 109 & 282\\  
2458053.855002 & 97 & 342\\   
2458054.907031 & 166 & 225\\  
2458055.960418 & 429 & 176\\  
2458056.910076 & 487 & 386\\  
2458060.896728 & -514 & 397\\ 
2458067.923497 & 172 & 372\\  
2458068.961460 & 85 & 157\\   
2458069.858188 & 143 & 346\\  
2458070.770534 & -433 & 271\\ 
2458071.786275 & 399 & 269\\  
2458099.839901 & -8 & 216\\   
2458099.862025 & 492 & 315\\  
2458099.884305 & 350 & 279\\  
2458099.906481 & 457 & 241\\  
2458106.891022 & -288 & 222\\
\midrule
\bottomrule
  \end{tabular}

\end{table}


For planets around rapidly rotating stars that are not conducive to a mass measurement via traditional radial velocity techniques, spectroscopic detections of planetary transits can be used to eliminate blend scenarios. During the transit, the planet casts a shadow on the rotating stellar surface, distorting the rotationally-broadened spectroscopic line profile that we observe. By obtaining a time series of spectroscopic observations during transit, we can track the changes in the line profiles, and detect the Doppler tomographic shadow cast by the planet \citep{Rossiter:1924,McLaughlin:1924,Cameron:2010}. Such a detection, when consistent with the amplitude expected based on the stellar \vsini and transit depth, as well as the impact parameter from the transit, confirms that the occultor is indeed transiting the primary star.  This, combined with an upper limit on the occultor's mass from the Doppler measurements, confirms the planetary nature of the occultor \citep{Cameron:2010,Bieryla:2015} 

We obtained a series of spectroscopic observations during the transit of \epic{} on 2018 Jan 04 with the Magellan Inamori Kyocera Echelle \citep[MIKE,][]{Bernstein:2003}, located at Las Campanas Observatory, Chile. MIKE is a slit-fed echelle spectrograph, with a wavelength coverage of $3200-10000$\,\AA{} obtained over the red and blue arms of the instrument. Our observations were obtained using the $0.\arcsec{}50\times5.\arcsec{}00$ slit, yielding a spectroscopic resolving power of $R=43,000$ in the red, and $R=48,000$ in the blue. A total of 29 observations were obtained through the transit, at an integration time of 600\,s in both the blue and red cameras. Wavelength solutions are provided by ThAr arc lamp exposures every 30\,minutes. Flat fielding corrections are applied using Quartz lamp exposures taken during the afternoon. The spectral reductions and extractions were performed using the Carnegie \emph{CarPy} package \citep{2000ApJ...531..159K,2003PASP..115..688K}.

The Doppler tomographic analysis largely follows that described in \citet{Zhou:2016a}. To measure the spectroscopic line profiles, we perform a least-squares deconvolution between the observed spectra and a non-rotating spectral template \citep{Donati:1997,Cameron:2010}. The resulting kernel from the deconvolution describes the line broadening profile, including effects from stellar rotation, macroturbulence, and instrumental broadening. The unbroadened infinite resolution spectral template is synthesised with the \emph{SPECTRUM} code \citep{Gray:1994}, using the ATLAS9 atmosphere models \citep{CastelliKurucz:2004}, at the measured stellar parameters of the host star. The deconvolution is performed over individual echelle orders spanning the spectral range $4900-6200$\,\AA{}, and the resulting broadening kernels from each order are averaged into a master broadening kernel for each observation.

\begin{figure}[ht] 
   \centering
   \includegraphics[width=0.5\textwidth]{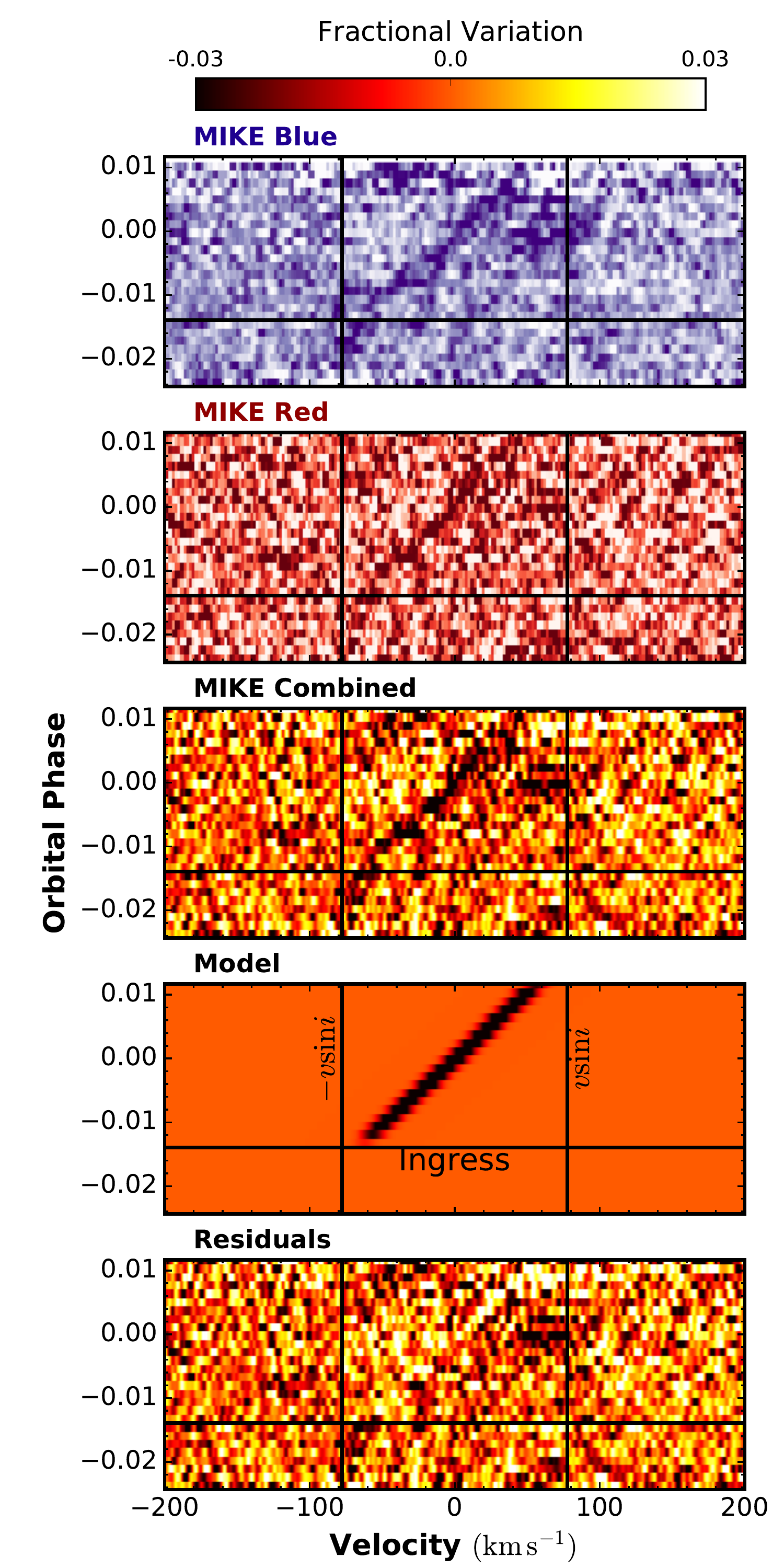} 
   \caption{Doppler tomographic map for the transit of \planet, as observed with MIKE/Magellan on 2018 Jan 04. The tomographic maps show the temporal variations of the stellar line profile as a function of orbital phase and stellar rotational velocity. The top two panels show the planetary signal as seen from the blue and red arms of the spectrograph. The middle panel shows the combined tomographic signal. The bottom two panels show the best fit transit model and the residuals after model subtraction.}
   \label{fig:DT}
\end{figure}

Changes in the line broadening kernel derived from the deconvolutions reveal the Doppler tomographic shadow of the planet. An averaged out-of-transit kernel is subtracted from each observation, and the resulting line profile residuals are plotted in the Doppler tomographic map in Fig.~\ref{fig:DT}. The dark trail represents the Doppler tomographic shadow of the transiting planet. The angle of the trail describes the projected spin-orbit angle of the planet, and is fitted for in the global model in Section~\ref{sec:exofast}.

\subsubsection{Keck/NIRC2 Adaptive Optics Imaging}\label{sec:ao}
As part of our standard process for validating transiting exoplanets, we observed EPIC~246851721 with infrared high-resolution adaptive optics (AO) imaging at Keck Observatory.  The Keck Observatory observations were made with the NIRC2 instrument on Keck-II behind the natural guide star AO system.  The observations were first made on 2017 Aug 20 in $Br-\gamma$ and the repeat follow-up observations were made on 2017 Oct 30 in both $Br-\gamma$ ($\lambda_\circ = 2.1686$ \micron, $\Delta\lambda = 0.0326$ \micron) and in the $J$ band ($\lambda_\circ = 1.248$ \micron, $\Delta\lambda = 0.163$ \micron) filters.  The observation sequence utilized the standard 3-point dither pattern that is used with NIRC2 to avoid the left lower quadrant of the detector which is typically noisier than the other three quadrants. The dither pattern step size was $3\arcsec$ and was repeated three times, with each dither offset from the previous dither by $0.5\arcsec$.  

\begin{figure*}[ht] 
  \centering
\plottwo{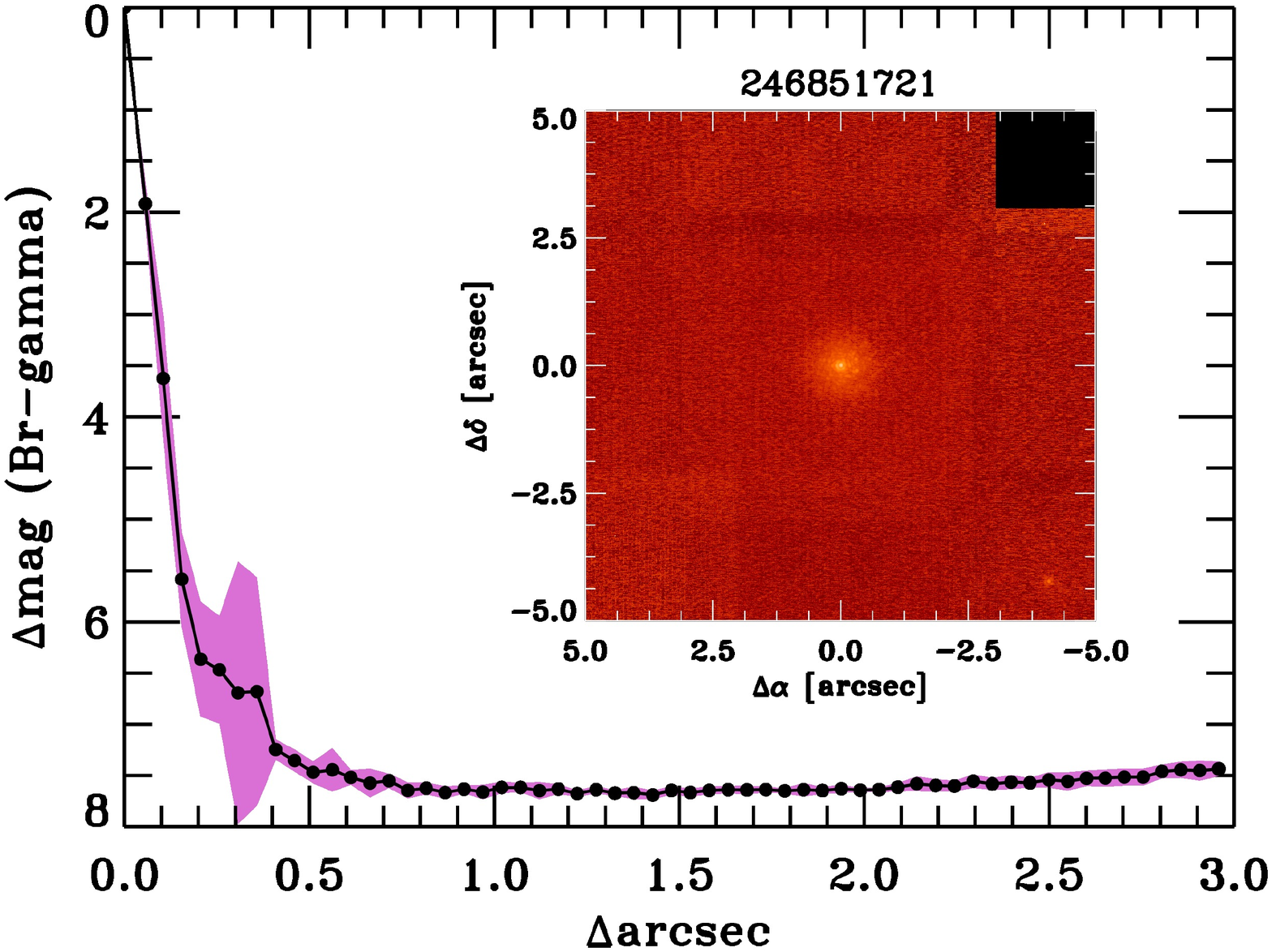}{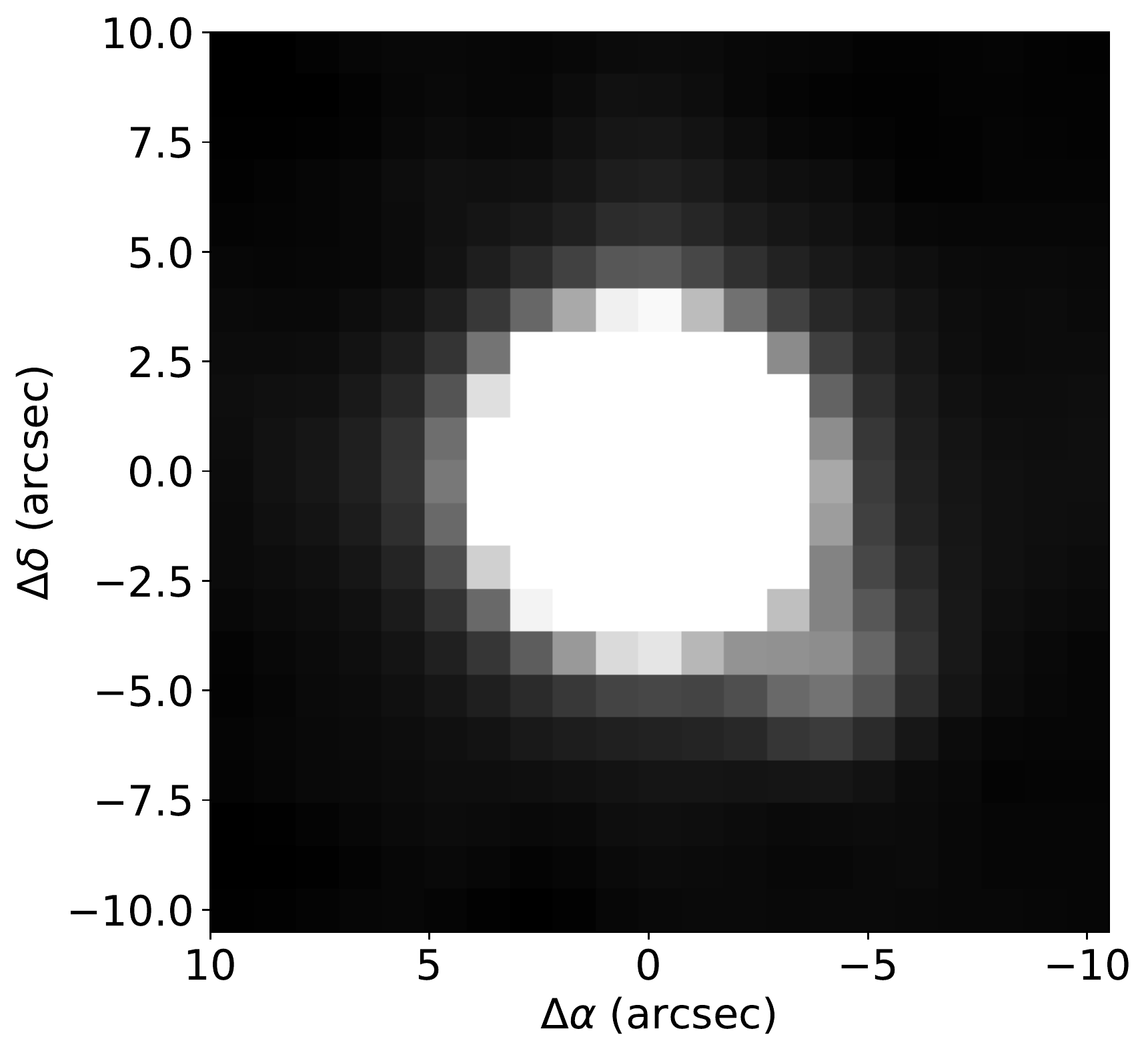}
   \caption{{\it Left:} AO images (inset) and $K_S$-band contrast curves for \epic, taken with Keck/NIRC2. To assess the effect of azimuthal structure on the sensitivity analysis, the image was divided into 45$^\circ$ ``pie-wedges" and the radial sensitivity was calculated in each of the eight wedges. The shading represents the rms azimuthal dispersion of the sensitivities for each of the radial steps. A companion is visible at 5.8$^{\prime\prime}$ southwest of the primary. {\it Right:} 2MASS $J$-band image showing \epic and its nearby companion.}
   \label{fig:ao}
\end{figure*}

The August 2018 observations utilized an integration time of 7.8 seconds with one coadd per frame for a total of 70.2 seconds and detected a companion 5.8\arcsec\ to the southwest (Fig.~\ref{fig:ao}).  The additional follow-up observations made in October 2018 utilized a 10 second integration (1 coadd) in the $Br-\gamma$ filter for a total of 90 seconds and a 1 second integration (1 coadd) in the $J$ band filter for a total of 9 seconds.  In all observations, the camera was in the narrow-angle mode with a full field of view of $10\arcsec$ and a pixel scale of approximately $0.009942\arcsec$ per pixel. We use the dithered images to remove sky background and dark current, and then align, flat-field, and stack the individual images. The NIRC2 AO data have a full-width half-maximum (FWHM) resolution of 0.051$^{\prime\prime}$ in $Br-\gamma$ and 0.039$^{\prime\prime}$ in the $J$ band. The sensitivities of the AO data were determined by injecting fake sources into the final combined images with separations from the primary targets in integer multiples of the central source's FWHM \citep{furlan2017, ciardi2018}. The companion star is $\sim4$ magnitudes fainter than the primary star in the $Ks$ band. 

The stellar companion at 5.8$^{\prime\prime}$ southwest of the primary star is also detected in the 2MASS imaging (2MASS~J05154048+1616394, see right panel of Fig.~\ref{fig:ao}).  In agreement with the Keck observations, the companion star is approximately 3.9 magnitudes fainter than the primary star at $Ks$ and 4.4 magnitudes fainter at $J$. 
Utilizing Kepler magnitude ($Kp$)-$Ks$ relationships from \citet{howell12}, we derive approximate deblended Kepler magnitudes of $Kp = 11.3$ mag for the primary and $Kp = 16.4$ mag for the companion. The resulting Kepler magnitude difference is $\Delta Kp = 5.1$ mag.

Doppler tomography shows that the transiting planet orbits the primary star and the AO imaging rules out the presence of any additional stars within $\sim 0.25$\arcsec\ of the primary and the presence of any additional brown dwarfs or widely-separated tertiary components beyond 0.25\arcsec. The presence of the blended 5.8\arcsec\ stellar companion is taken into account to obtain the correct transit depth and planetary radius \citep{ciardi2015}.

\begin{figure*}[htb]
\centering
\includegraphics[width=\textwidth, trim={0cm 7cm 0cm 7cm}, clip]{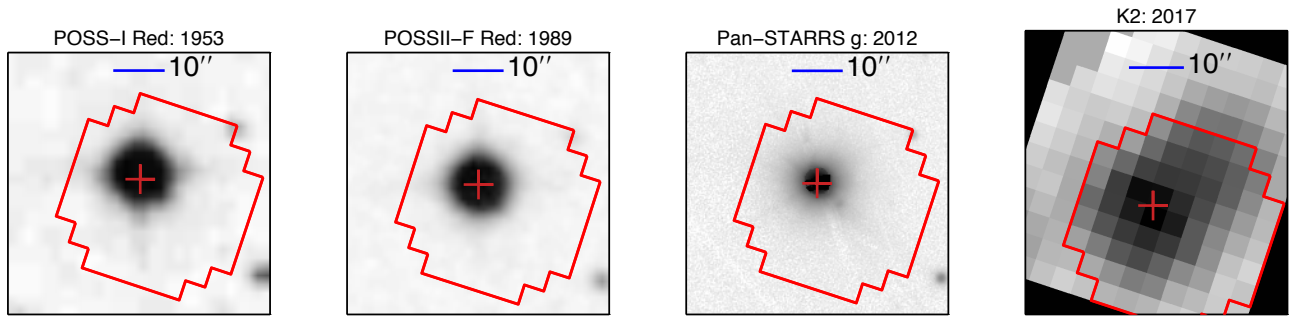}
\caption{Left to right: POSS-I red band image showing \epic, with the \ktwo aperture overplotted in red; same, but for POSS-II red band; same, but for Pan-STARRS $g$ band; \ktwo pixel-level image from which we extracted the \ktwo photometry, with the aperture overplotted in red.}
\label{fig:ap}
\end{figure*}

In addition, we also examined archival images from POSS-I, POSS-II and Pan-STARRS (Fig.~\ref{fig:ap}). POSS-I and Pan-STARRS both show another possible faint source northeast of the primary, right at the edge of the \ktwo aperture, but the corresponding catalogs as well as UCAC, 2MASS, SDSS12 and GSC2.3 all indicate that there are no other sources brighter than 20 magnitudes in the $g'$-band and 19 magnitudes in the $r'$-band within 30$^{\prime\prime}$ of the primary, apart from the companion at 5.8\arcsec. The optical flux contribution of any other possible nearby sources is below the precision of \ktwo and can be safely ignored in our transit fits.

\subsubsection{Ground-Based Photometry}\label{demonext}
We obtained additional ground-based photometry for \epic using the DEMONEXT telescope \citep{demonext} at Winer Observatory in Sonoita, Arizona. DEMONEXT is a 0.5 m PlaneWave CDK20 $f$/6.8 Corrected Dall-Kirkham Astrograph telescope. It has a 2048$\times$2048 pixel FLI Proline CCD3041 camera, with a $30.7' \times 30.7'$ field of view and a pixel scale of $0.90^{\prime\prime}$ pixel$^{-1}$.

\epic was placed in the DEMONEXT automated queue beginning in December 2017 through February 2018 with observations requested in $i'$. Using the ephemerides derived from the \ktwo light curves, RA, DEC, and magnitude, DEMONEXT automatically scheduled transits on 5 nights where at least 1.5 hours of observations could be made that included 30 minutes of either pre-ingress or post-egress data, and 60 minutes of in-transit data. Observations ranged from 2 to 6 hours. An exposure time of 20 seconds was used and DEMONEXT was defocused to avoid saturation. In February 2018, DEMONEXT was also used in alternating filter mode with observations alternating between $g'$ and $i'$ for the transit occurring on UT 2018 Feb 05. A total of 1882 observations were made, with 1769 in $i'$ and 113 in $g'$. 

\begin{figure}
    \centering
    \includegraphics[width=0.45\textwidth]{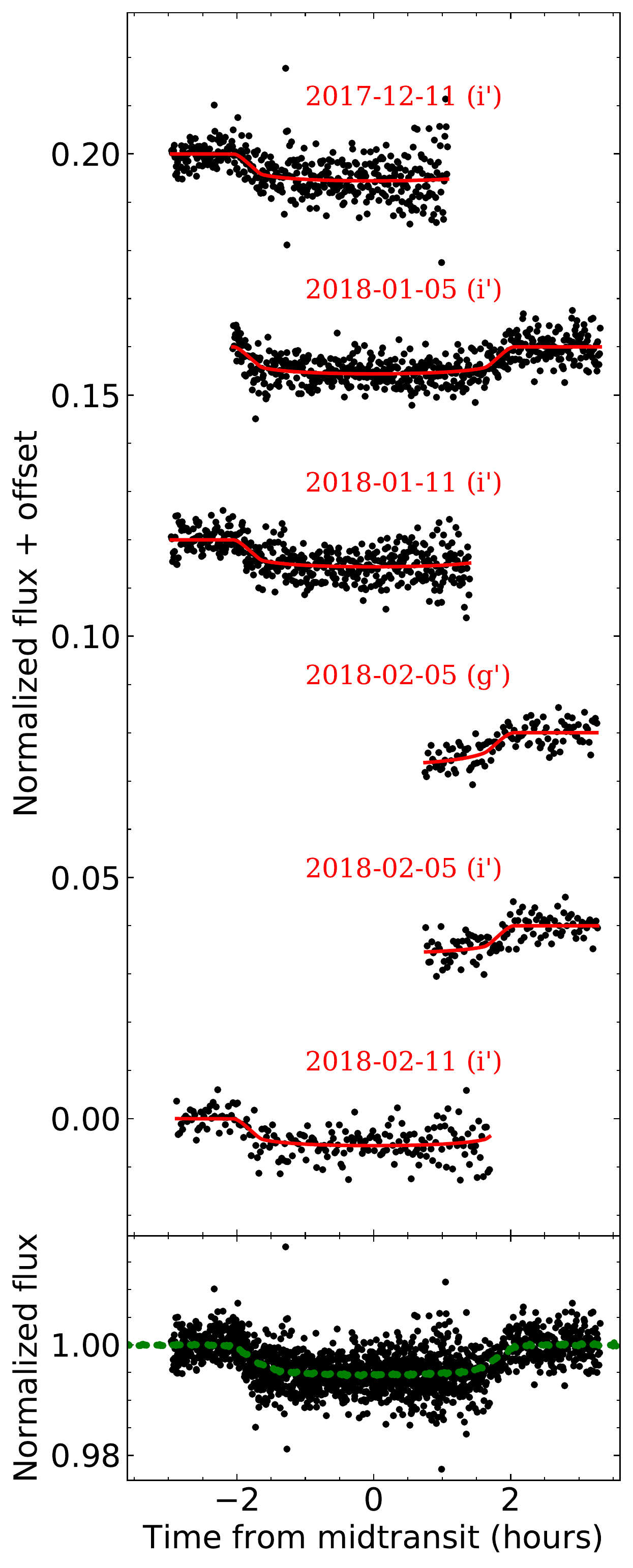}
    \caption{Top: Individual transit observations of \epic from DEMONEXT with the best-fit model from the global fit (Section~\ref{sec:exofast}) overplotted in red. A vertical offset is applied to each light curve for clarity. Bottom: The phase-folded DEMONEXT light curves, with the phase-folded \ktwo light curve overplotted in green.}
    \label{fig:demonext}
\end{figure}

All observations were reduced using standard bias, dark, and flat-fielding techniques. Relative aperture photometry was performed using AIJ \citep{collins2017} on the defocused images to obtain the time-series light curves. No detrending parameters were used in the initial reductions. The resulting light curves are shown in Fig.~\ref{fig:demonext}. The stellar companion at 5.8$^{\prime\prime}$ southwest of the primary star is not resolved in the defocused images.

\section{Host Star Characterization}\label{star}
\subsection{Spectral Analysis}
We obtain initial estimates of some of \epic's physical properties from the TRES spectra using the Spectral Parameter Classification (SPC) procedure of \citet{buchhave12}. However, since the star is rapidly rotating, SPC yields less reliable results than for slowly rotating stars. We ran SPC with no parameters fixed, and obtain the following error-weighted mean values: effective temperature \teff $=6565 \pm 203$ K, surface gravity $\log g=3.92 \pm 0.44$ (cgs) and metallicity $[m/H]=0.14 \pm 0.09$.
We adopt these values as starting points and/or priors for the global fit described in Section~\ref{sec:exofast}.

Since an accurate measurement of the projected stellar rotational velocity \vsini directly affects the measured spin-orbit angle of the planet, we performed a detailed modelling of the line profile to derive the broadening velocities of the host star. We derive a series of line broadening kernels from the TRES spectra via a least-squares deconvolution against synthetic non-rotating templates (similar to the process described in Section~\ref{spectroscopy}). We make use of the TRES spectra since TRES is a fibre-fed spectrograph with a stable instrument profile. We model the line broadening profiles via a numerical disk integration, accounting for the varying radial-tangential macroturbulence, rotational broadening, and limb darkening in our model \citep[following][]{Graybook:2005}. The disk-integrated line broadening kernel is then convolved with a Gaussian function of width $6.8\,\mathrm{km\,s}^{-1}$ to account for the instrument broadening. The disk integration process is computationally intensive, as such we only compute a grid of broadening kernels spanning $2 < v\sin I_\star < 100\,\mathrm{km\,s}^{-1}$ and $1 < v_\mathrm{macro} < 20\,\mathrm{km\,s}^{-1}$, and interpolate within the grid during our model fit \citep{Zhou18}. By modeling the line profiles, we derive a projected stellar rotation rate of $v\sin I_\star = 74.77 \pm 0.63\,\mathrm{km\,s}^{-1}$ , and a macroturbulent broadening of $v_\mathrm{macro} = 6.79 \pm 0.62 \,\mathrm{km\,s}^{-1}$. The uncertainties are determined by the scatter of the broadening velocities from observation to observation, and may be underestimated as they do not account for systematic offsets due to model mismatch. These broadening velocities are used in the global modeling of the system described in Section~\ref{sec:exofast}.

\subsection{UVW Space Motion}
We calculate the star's Galactic space-velocity components $U$, $V$ and $W$ to locate it kinematically within the Galaxy.  We use the Gaia DR2 parallax \citep{gaia18} (uncorrected for the \citet{stassun2018} systematic offset of 80 $\mu$arcseconds), to infer a distance
to EPIC 246851721 of $375\pm18~{\rm pc}$.  Note that the \citet{stassun2018} offset is slightly smaller than the DR2-quoted uncertainty of this star of 130~$\mu$arcseconds.  We find that EPIC 246851721 is located about 80 parsecs below the plane, which is entirely typical for a thin disk star of this spectral type \citep{bovy2017}.  Using the absolute systemic RV of this system from TRES of $36.85^{+0.13}_{-0.12}$ \kms, we infer that $(U,V,W)$=($-22.6 \pm 0.3, -12.6 \pm 1.1, -10.1\pm 0.5$)~$~{\rm km~s^{-1}}$; again entirely normal kinematics for a star of this spectral type and confirming that it is, indeed, a thin disk star \citep{Bensby:2003}.

Had we applied the \citet{stassun2018} offset, we would have inferred a distance of $373\pm 18$~pc, well within $1\sigma$ of the uncorrected result;  doing so would have had a negligible impact on our inferences about the location and kinematics of this star.

\section{EXOFASTv2 Global Fit}\label{sec:exofast} 
To properly determine the final system parameters for \epic and its planet, we use the exoplanet global fitting suite EXOFASTv2 \citep{Eastman:2017}, to perform a global analysis of all the available data. Specifically, we perform a simultaneous fit of the flattened {\it K2} light curve, accounting for the long cadence smearing; the DEMONEXT light curves; the Doppler tomographic map; and the radial velocity data. Within the fit, we use the spectral energy distribution (SED), the integrated YY stellar tracks \citep{Yi:2001}, and the parallax from Gaia DR2 summarized in Table \ref{tab:stellar} to constrain the host star properties with priors set on \teff and \feh \ from the TRES spectra described in Section~\ref{spectroscopy}. We enforce an upper limit on the V-band extinction from the \citet{Schlegel1998} dust maps of 0.328. Since the high \vsini is causing higher than typical errors on our radial velocity measurements from TRES, we do not attempt to fit for the eccentricity and fix it to zero. To correct for the flux contribution from the nearby blended stellar companion, we separately fit its SED and determine the following deblending coefficients (B/A) for the \ktwo and DEMONEXT light curves: Kepler = 0.001597, $g^\prime$ = 0.005259, $i^\prime$ = 0.000792. We include these deblending coefficients in our EXOFASTv2 global fit to properly account for the contamination on each light curve. We note that our final determined parallax from the global fit is 2.970$\pm$0.072 mas, which is $\sim2\sigma$ different from the value reported by Gaia (2.66$\pm$0.13 mas). It is not clear what is causing this discrepancy, but the higher than typical error on the Gaia parallax and the blending nearby companion may be responsible. 

\begin{figure*}[htb]
\centerline{%
\includegraphics[width=0.5\textwidth, trim={2cm 12.5cm 8cm 8cm}, clip]{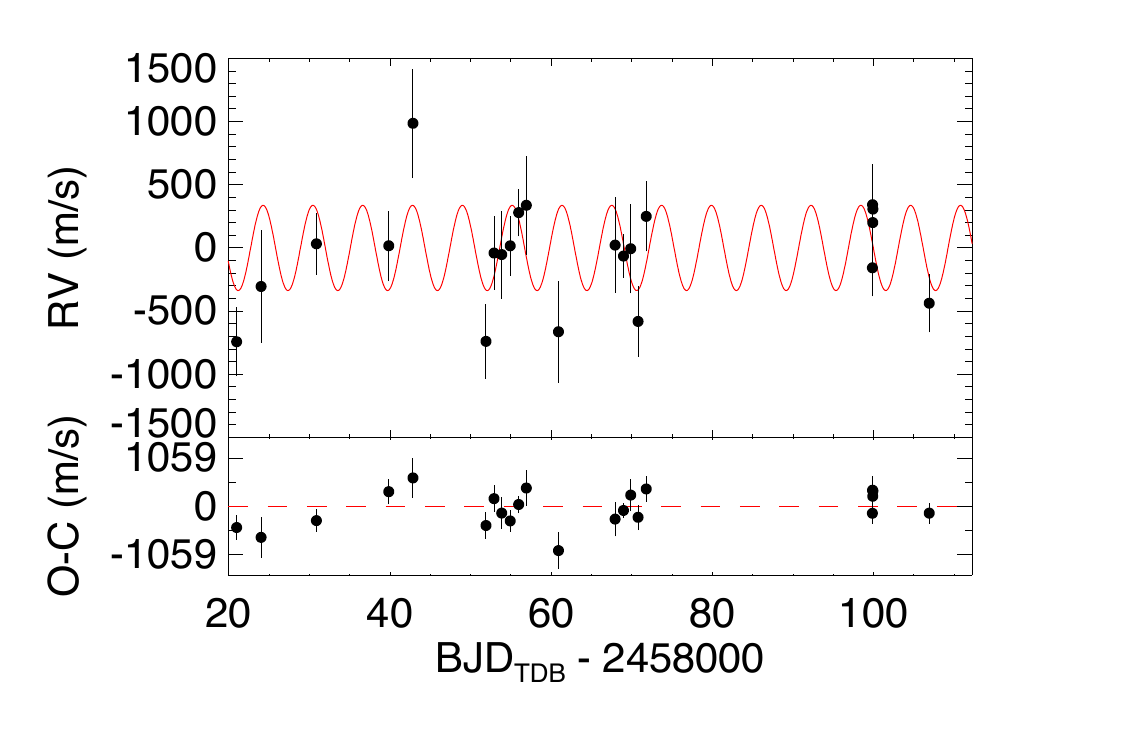}%
\includegraphics[width=0.5\textwidth, trim={2cm 12.5cm 8cm 8cm}, clip]{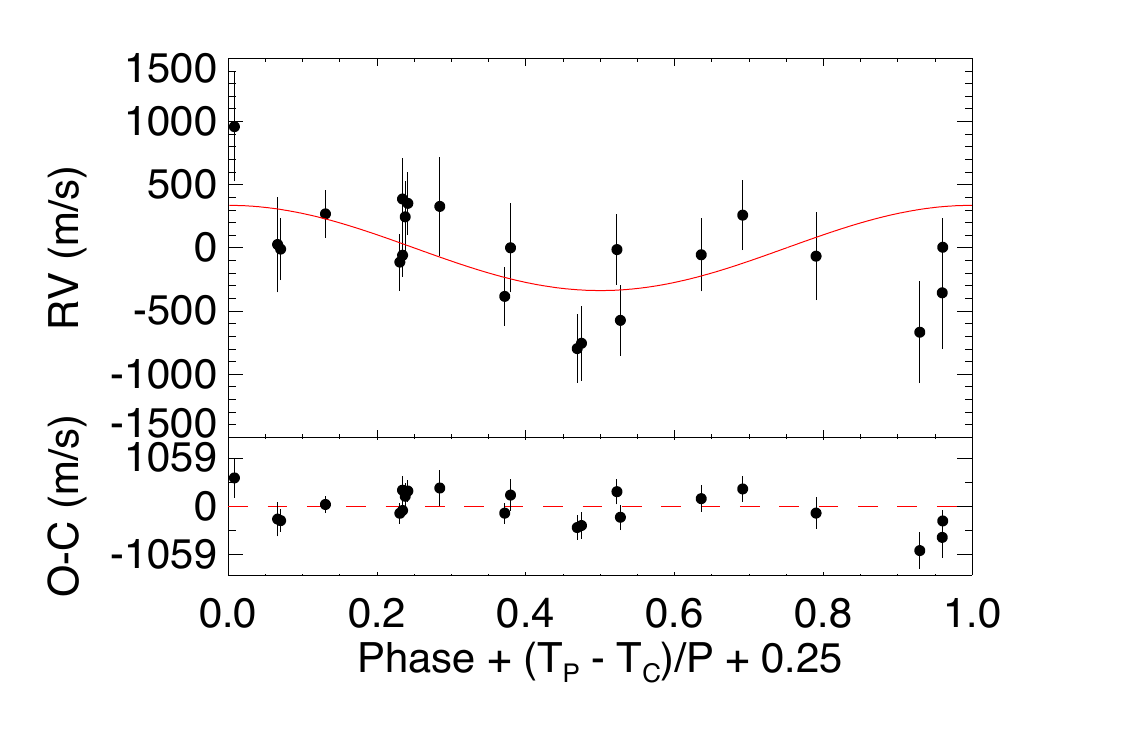}%
}%
\caption{{\it Left}: The RV time series of \epic from TRES. The maximum-likelihood Keplerian orbital model is plotted in red. The instrumental offset has been subtracted from each data set and the model. The uncertainties plotted include the RV jitter terms listed in Table~\ref{tab:params} added in quadrature with the measurement uncertainties for all RVs. Below are the residuals to the maximum-likelihood orbit model. {\it Right}: same as the left panel, but phase-folded to the best-fit ephemeris. The X-axis is defined such that the primary transit occurs at 0.25, where $T_P$ is the time of periastron, $T_C$ is the time of transit, and $P$ is the period.}
\label{fig:rv} 
\end{figure*}

\begin{figure}[htb]
    \centering
    \includegraphics[width=0.5\textwidth]{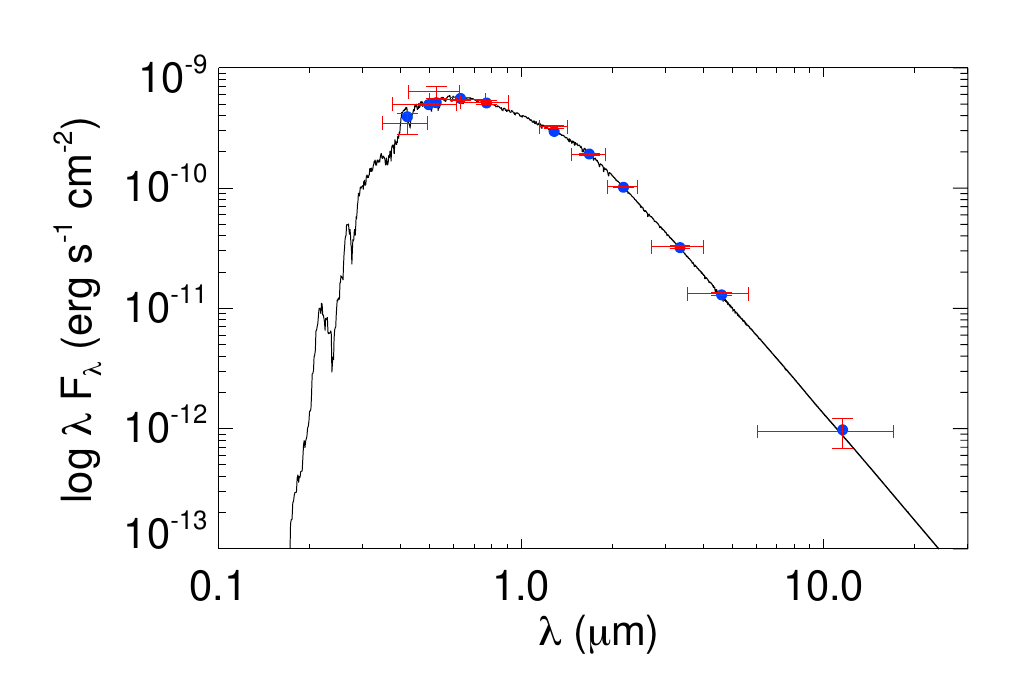}
    \caption{SED fits to \epic from EXOFASTv2. The red points show observed values, with the vertical error bars representing $1\sigma$ measurement uncertainties and horizontal error bars representing the widths of the bandpasses. The blue points are the model fluxes in the observed bandpasses. The solid lines show the model fits.}
    \label{fig:sed}
\end{figure}

\begin{figure}[htb]
    \centering
    \includegraphics[width=0.5\textwidth, trim={2cm 2cm 2cm 2cm}, clip]{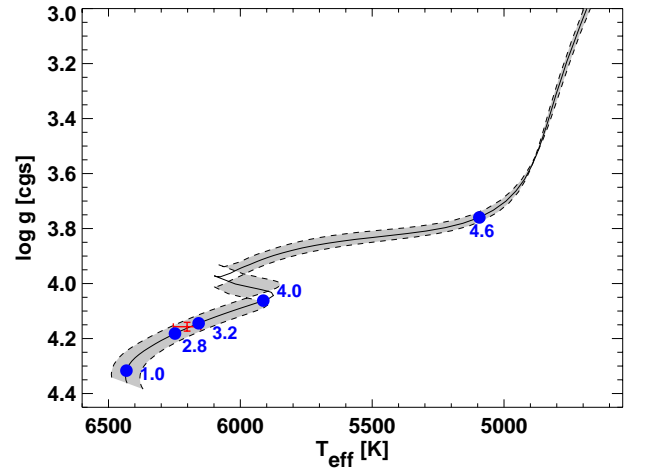}
    \caption{The location of \epic in the Kiel diagram. The median \teff and \logg from the global model fit are shown as the red point, while the grey swath shows the YY evolutionary track for a star with the best-fit values of $M_\mathrm{\star}$ and [Fe/H]; the locations on the best-fit model corresponding to several values of stellar age are shown as blue points, with ages quoted in Gyr.}
    \label{fig:yy}
\end{figure}

See Fig.~\ref{fig:lc} for the final {\it K2} transit fit, Fig.~\ref{fig:rv} for the final RV fit, Fig.~\ref{fig:sed} for the final SED fit from our EXOFASTv2 global fit, and Fig.~\ref{fig:yy} for the best-fit evolutionary track. The median values of the posterior distributions of the system parameters are shown in Table \ref{tab:params}, and the telescope parameters are shown in Table \ref{tab:params2}. 

\begin{table*}
\scriptsize
\centering
\setlength\tabcolsep{1.5pt}
\caption{Median values and 68\% confidence interval for \epic}
\label{tab:params}
\begin{tabular}{lcccc}
  \hline
  \hline
~~~Parameter &  Units & \multicolumn{3}{c}{Values}\\

\hline
Stellar Parameters: & & & & \\
\hline
~~~~$M_*$\dotfill &Mass (\msun)\dotfill &$1.317^{+0.041}_{-0.040}$\\
~~~~$R_*$\dotfill &Radius (\rsun)\dotfill &$1.586^{+0.034}_{-0.033}$\\
~~~~$L_*$\dotfill &Luminosity (\lsun)\dotfill &$3.36^{+0.20}_{-0.18}$\\
~~~~$\rho_*$\dotfill &Density (cgs)\dotfill &$0.465^{+0.025}_{-0.024}$\\
~~~~$\log{g}$\dotfill &Surface gravity (cgs)\dotfill &$4.157\pm0.016$\\
~~~~$T_{\rm eff}$\dotfill &Effective Temperature (K)\dotfill &$6202^{+52}_{-50}$\\
~~~~$[{\rm Fe/H}]$\dotfill &Metallicity \dotfill &$0.141^{+0.082}_{-0.085}$\\
~~~~$Age$\dotfill &Age (Gyr)\dotfill &$3.02^{+0.44}_{-0.46}$\\
~~~~$vsinI_*$\dotfill &Projected rotational velocity (km/s)\dotfill &$74.92^{+0.62}_{-0.60}$\\
~~~~$macturb$\dotfill &Macroturbulence (m/s)\dotfill &$7220\pm590$\\
~~~~$A_v$\dotfill &V-band extinction \dotfill &$0.315^{+0.014}_{-0.030}$\\
~~~~$\sigma_{SED}$\dotfill &SED photometry error scaling \dotfill &$1.21^{+0.49}_{-0.29}$\\
~~~~$d$\dotfill &Distance (pc)\dotfill &$336.7^{+8.4}_{-8.0}$\\
~~~~$\pi$\dotfill &Parallax (mas)\dotfill &$2.970\pm0.072$\\
~~~~$\dot{\gamma}$\dotfill &RV slope (m/s/day)\dotfill &$2.8^{+3.0}_{-2.9}$\\
\hline
Planetary Parameters:\\
\hline
~~~~$P$\dotfill &Period (days)\dotfill &$6.180235\pm0.000014$\\
~~~~$R_P$\dotfill &Radius (\rj)\dotfill &$1.051\pm0.044$\\
~~~~$T_C$\dotfill &Time of conjunction (\bjdtdb)\dotfill &$2457858.93009\pm0.00010$\\
~~~~$T_0$\dotfill &Optimal conjunction Time (\bjdtdb)\dotfill &$2457865.11032\pm0.00010$\\
~~~~$a$\dotfill &Semi-major axis (AU)\dotfill &$0.07229^{+0.00074}_{-0.00075}$\\
~~~~$i$\dotfill &Inclination (Degrees)\dotfill &$86.21\pm0.17$\\
~~~~$T_{eq}$\dotfill &Equilibrium temperature (K)\dotfill &$1401\pm16$\\
~~~~$M_P$\dotfill &Mass (\mj)\dotfill &$3.0^{+1.1}_{-1.2}$\\
~~~~$K$\dotfill &RV semi-amplitude (m/s)\dotfill &$270^{+100}_{-110}$\\
~~~~$logK$\dotfill &Log of RV semi-amplitude \dotfill &$2.44^{+0.14}_{-0.21}$\\
~~~~$R_P/R_*$\dotfill &Radius of planet in stellar radii \dotfill &$0.0681^{+0.0025}_{-0.0026}$\\
~~~~$a/R_*$\dotfill &Semi-major axis in stellar radii \dotfill &$9.80\pm0.17$\\
~~~~$\delta$\dotfill &Transit depth (fraction)\dotfill &$0.00464\pm0.00035$\\
~~~~$Depth$\dotfill &Flux decrement at mid transit \dotfill &$0.00464\pm0.00035$\\
~~~~$\tau$\dotfill &Ingress/egress transit duration (days)\dotfill &$0.01810^{+0.00092}_{-0.00093}$\\
~~~~$T_{14}$\dotfill &Total transit duration (days)\dotfill &$0.17117^{+0.00074}_{-0.00075}$\\
~~~~$T_{FWHM}$\dotfill &FWHM transit duration (days)\dotfill &$0.15307^{+0.00052}_{-0.00050}$\\
~~~~$b$\dotfill &Transit Impact parameter \dotfill &$0.647^{+0.017}_{-0.018}$\\
~~~~$\delta_{S,3.6\mu m}$\dotfill &Blackbody eclipse depth at 3.6$\mu$m (ppm)\dotfill &$251\pm20$\\
~~~~$\delta_{S,4.5\mu m}$\dotfill &Blackbody eclipse depth at 4.5$\mu$m (ppm)\dotfill &$355\pm27$\\
~~~~$\rho_P$\dotfill &Density (cgs)\dotfill &$3.1\pm1.3$\\
~~~~$logg_P$\dotfill &Surface gravity \dotfill &$3.82^{+0.14}_{-0.21}$\\
~~~~$\lambda$\dotfill &Projected Spin-orbit alignment (Degrees)\dotfill &$-1.47^{+0.87}_{-0.86}$\\
~~~~$\Theta$\dotfill &Safronov Number \dotfill &$0.31\pm0.12$\\
~~~~$\fave$\dotfill &Incident Flux (\fluxcgs)\dotfill &$0.874^{+0.041}_{-0.039}$\\
~~~~$T_P$\dotfill &Time of Periastron (\bjdtdb)\dotfill &$2457858.93009\pm0.00010$\\
~~~~$T_S$\dotfill &Time of eclipse (\bjdtdb)\dotfill &$2457862.02021\pm0.00010$\\
~~~~$T_A$\dotfill &Time of Ascending Node (\bjdtdb)\dotfill &$2457857.38503\pm0.00010$\\
~~~~$T_D$\dotfill &Time of Descending Node (\bjdtdb)\dotfill &$2457860.47515\pm0.00010$\\
~~~~$M_P\sin i$\dotfill &Minimum mass (\mj)\dotfill &$3.0^{+1.1}_{-1.2}$\\
~~~~$M_P/M_*$\dotfill &Mass ratio \dotfill &$0.00215^{+0.00081}_{-0.00086}$\\
~~~~$d/R_*$\dotfill &Separation at mid transit \dotfill &$9.80\pm0.17$\\
~~~~$P_T$\dotfill &A priori non-grazing transit prob \dotfill &$0.0951\pm0.0017$\\
~~~~$P_{T,G}$\dotfill &A priori transit prob \dotfill &$0.1090\pm0.0019$\\
\hline
\multicolumn{2}{l}{Wavelength Parameters:}&Kepler&g'&i' \\
\hline
~~~~$u_{1}$\dotfill &linear limb-darkening coeff \dotfill &$0.250\pm0.031$&$0.500^{+0.051}_{-0.050}$&$0.228\pm0.023$\\
~~~~$u_{2}$\dotfill &quadratic limb-darkening coeff \dotfill &$0.270\pm0.042$&$0.267\pm0.050$&$0.291\pm0.022$\\
~~~~$A_D$\dotfill &Dilution from neighboring stars \dotfill &$-0.109^{+0.078}_{-0.089}$&$-0.22^{+0.14}_{-0.16}$&$-0.134^{+0.088}_{-0.100}$\\
\hline
\multicolumn{2}{l}{Telescope Parameters:}&TRES \\
\hline
~~~~$\gamma_{\rm rel}$\dotfill &Relative RV Offset (m/s)\dotfill &$90^{+73}_{-71}$\\
~~~~$\sigma_J$\dotfill &RV Jitter (m/s)\dotfill &$200^{+110}_{-130}$\\
~~~~$\sigma_J^2$\dotfill &RV Jitter Variance \dotfill &$40000^{+58000}_{-34000}$\\
\hline
 \end{tabular}
\end{table*}

\begin{table*}
 \scriptsize
 \centering
 \setlength\tabcolsep{1.5pt}
 \caption{Median values and 68\% confidence interval for the Telescope parameters of \epic}
 \label{tab:params2}
 \begin{tabular}{lcccc}
   \hline
   \hline
 \multicolumn{2}{l}{Transit Parameters:}\\
 ~~~Observation &  Added Variance & Baseline flux & Additive detrending coeff\\
 ~~~~&$\sigma^{2}$ &$F_0$ &$C_{0}$ \\
 UT 2017 K2 C13 (Kepler) &$-1.229^{+0.060}_{-0.055} \times 10^{-8}$&$1.0000013\pm0.0000048$&--\\
 DEMONEXT UT 2017-12-11 (i') &$8.8^{+1.1}_{-1.0}\times 10^{-6}$&$1.00373^{+0.00022}_{-0.00023}$&$0.00038\pm0.00052$\\
 DEMONEXT UT 2018-01-05 (i') &$1.39^{+0.43}_{-0.39}\times 10^{-6}$&$1.00371\pm0.00017$&$0.00104^{+0.00036}_{-0.00037}$\\
 DEMONEXT UT 2018-01-11 (i') &$3.50^{+0.63}_{-0.58}\times 10^{-6}$&$1.00388\pm0.00019$&$0.00025\pm0.00042$\\
 DEMONEXT UT 2018-02-05 (g') &$1.14^{+0.77}_{-0.65}\times 10^{-6}$&$1.00207^{+0.00030}_{-0.00029}$&$0.00144\pm0.00067$\\
 DEMONEXT UT 2018-02-05 (i') &$1.9^{+8.8}_{-7.4}\times 10^{-7}$&$1.00205\pm0.00024$&$0.00057^{+0.00066}_{-0.00065}$\\
 DEMONEXT UT 2018-02-11 (i') &$2.91^{+0.95}_{-0.84}\times 10^{-7}$&$1.00389\pm0.00025$&$-0.00118^{+0.00061}_{-0.00062}$\\
   \hline
   \hline
  \end{tabular}
 \end{table*}

\section{Discussion}
\label{discussion}

\subsection{The Stellar Companion}
The stellar companion to \epic has a parallax of $2.70 \pm 0.31$ mas and proper motions of $\mu_{\alpha}=5.69 \pm 0.48$ mas yr$^{-1}$ and $\mu_{\delta}= -14.81 \pm 0.35$ mas yr$^{-1}$ \citep{gaia18}. The parallax places the companion at a distance of $370 \pm 42$ pc, consistent with the distance of the primary within $1\sigma$. Their proper motions are also sufficiently similar that we can conclude that the companion is most likely bound. The projected separation of $5.842\arcsec$ translates to a physical separation of $\sim 2100$ AU. This places the system just beyond the range of separations examined by the Friends of Hot Jupiters collaboration in a series of papers \citep[][and references therein]{ngo16}, which found that stars hosting hot Jupiters are more likely to have stellar companions between 50 and 2000 AU compared to field stars. \planet therefore adds to the sample of hot and tropical Jupiters with bound stellar companions.

However, the properties of hot Jupiters' stellar companions are incompatible with high eccentricity migration through stellar Kozai-Lidov oscillations being the dominant channel of giant planet migration. \citet{ngo16} determined that only $16\% \pm 5\%$ of hot Jupiters have stellar companions capable of inducing Kozai-Lidov oscillations, assuming initial semi-major axes between 1 and 5 AU. \citet{dawson13} also showed that giant planets orbiting metal-poor stars are confined to lower eccentricities than those orbiting metal-rich stars. This is more compatible with planet-planet interactions than with planet-stellar Kozai-Lidov interactions, because giant planet formation is strongly correlated with stellar metallicity but stellar multiplicity is not. 

Nonetheless, we can explore the feasibility of stellar Kozai-Lidov emplacement in this system. In order to excite Kozai-Lidov oscillations on the planet, the precession caused by the perturber must be able to overcome other perturbing forces, including due to general relativity (GR), tides, and the oblateness of the star \citep{fabrycky10}. For Jupiter-sized planets with periods $\gtrsim$ 3 days, apsidal precession due to GR typically dominates over other sources of precession, so the Kozai-Lidov oscillation timescale must be shorter than the GR precession timescale in order for eccentricity excitation to take place. The former is on the order of \citep{kiseleva98}
\begin{equation}
\label{eq:kl}
    t_{KL} = \frac{2P_\mathrm{per}^2}{3\pi P} \frac{M_* + M_P + M_\mathrm{per}}{M_\mathrm{per}} (1-e^2_\mathrm{per})^{3/2}
\end{equation}
and the GR precession rate can be estimated as \citep[Eq. 23 from][]{fabryckytremaine}
\begin{equation}
\label{eq:gr}
    \dot{\omega}_\mathrm{GR} = \frac{3G^{3/2}(M_*+M_P)^{3/2}}{a_P^{5/2}c^2(1-e_P^2)}
\end{equation}
where $P$, $a_P$ and $e_P$ are the planet's initial orbital period, semi-major axis and eccentricity, and $M_\mathrm{per}$, $P_\mathrm{per}$ and $e_\mathrm{per}$ are the stellar perturber's mass, orbital period and eccentricity respectively. The inverse of this rate gives an estimate of the GR precession timescale. 

Using broadband photometry from 2MASS and SDSS, and Gaia parallaxes, we fit for the mass of the companion, $M_\mathrm{per}$, with the \texttt{isochrones} package\footnote{https://github.com/timothydmorton/isochrones} \citep{isochrones}. The companion appears to be an M dwarf with a mass of $\sim 0.4$ \msun. Without eccentricity constraints for the planet and the stellar companion, we assume a circular orbit for the planet and an orbital eccentricity of 0.5 for the companion, following \citet{ngo16}. Substituting these values and the sky-projected separation of 2100 AU into Eqs.~\ref{eq:kl} and~\ref{eq:gr}, we find that if the planet had a starting semi-major axis of 1 AU, the GR timescale is three orders of magnitude smaller than the Kozai-Lidov timescale and stellar Kozai-Lidov emplacement is therefore impossible. At a starting semi-major axis of 5 AU, the two timescales are comparable so that stellar Kozai-Lidov emplacement is in theory plausible. Yet if stellar Kozai-Lidov oscillations were really responsible for the migration of \planet, it seems unlikely that the system should have such a low obliquity (see Section~\ref{obliquity}). These mysteries can be solved with a larger sample of hot Jupiters with ``friends" and well-constrained obliquities.


\subsection{Spin-Orbit Alignment}\label{obliquity}
\begin{figure}[tb] 
   \centering
   \includegraphics[width=0.5\textwidth]{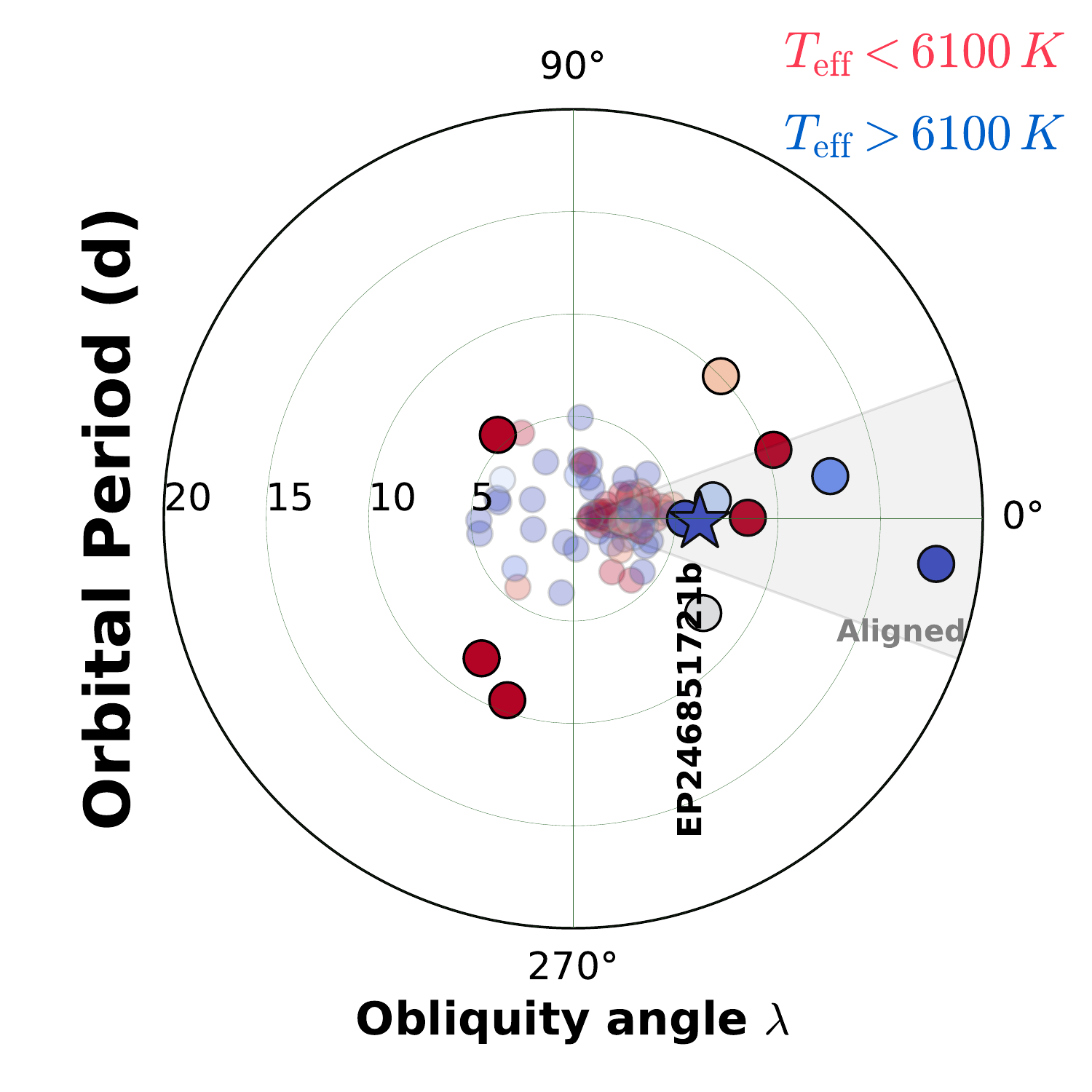} 
   \caption{Polar plot showing all planets with known obliquities. The angle shows the sky-projected obliquity $\lambda$ while the radius corresponds to the orbital period of each system. The colors scale with the \teff of the host stars. Planets with $P \ge 5$ days are shown in bold. Figure format inspired by J. Winn.}
   \label{fig:lambda}
\end{figure}

Doppler tomography also enabled us to measure the sky-projected spin-orbit misalignment $\lambda$ for \epic. Our global fit reveals that \planet has a very small projected obliquity of $-1.48^\circ \pm 0.85^\circ$. Fig.~\ref{fig:lambda} shows \planet alongside all other planets with known obliquities.  \planet joins a small group of known Jovian-sized planets at periods longer than 5 days (corresponding to $a/R_* \gtrsim 10$) orbiting stars near or above the Kraft break (\teff $\gtrsim$ 6100 K). Among these, only seven systems have measured projected obliquities. The other six are HAT-P-2 b \citep{winn07}, HAT-P-34 b \citep{albrecht12}, KELT-6 b \citep{damasso15}, KOI-12 b \citep{bourrier15}, KOI-94 d \citep{albrecht13} and WASP-38 b \citep{brown12}. \epic is the fourth hottest star in the group, and Fig.~\ref{fig:yy} suggests it has been even hotter in the past. The angle $\lambda$ is only the sky-projected angle between the stellar spin and orbital angular momentum vectors. To date, very few planets have known three-dimensional spin-orbit angles $\psi$, which cannot be directly measured. We can attempt to derive the true three-dimensional obliquity for this system using the information we have. 

\begin{figure}[hb] 
   \centering
   \includegraphics[width=0.45\textwidth]{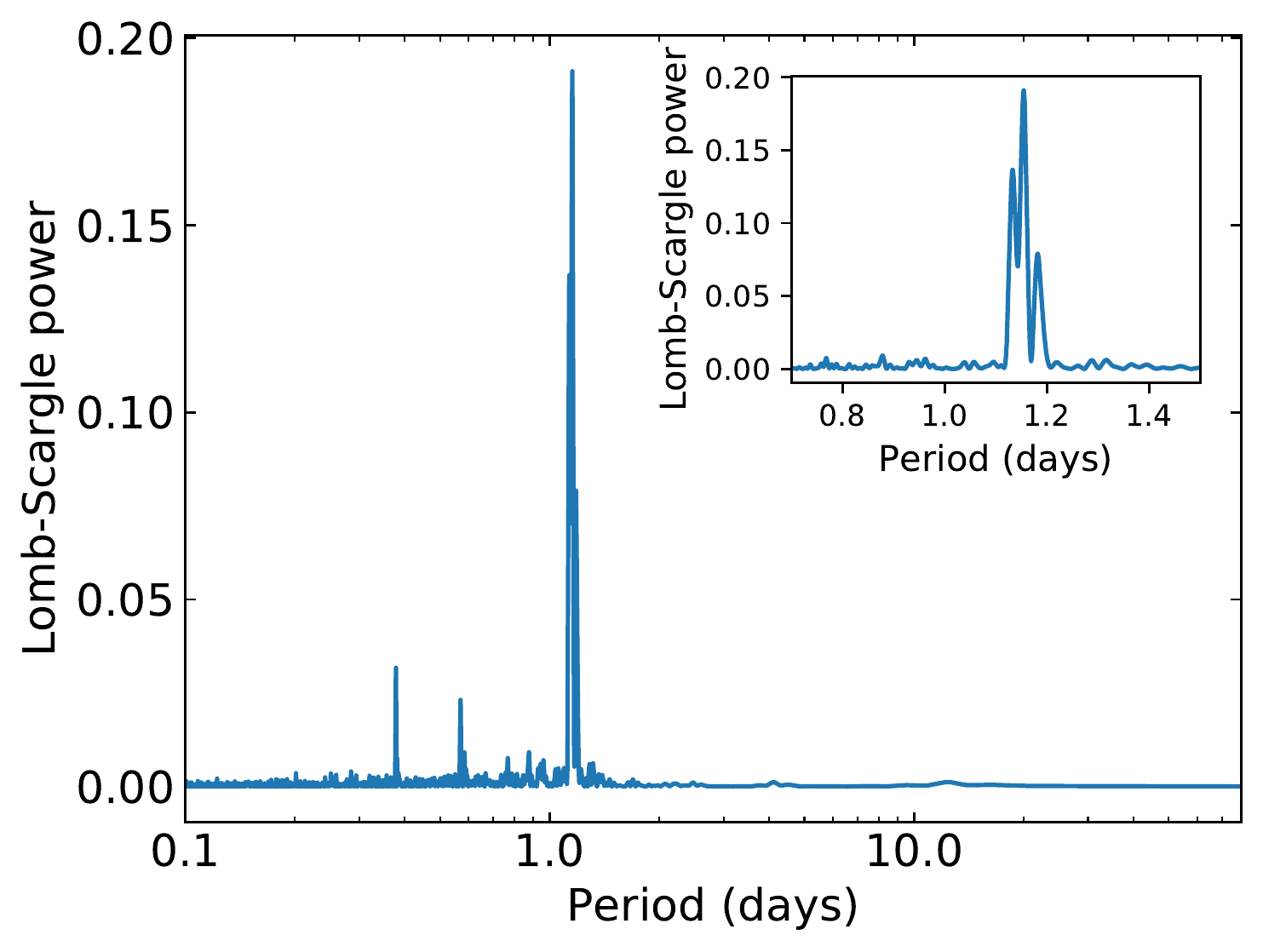} 
   \caption{Lomb-Scargle periodogram of the \ktwo light curve, with transits masked out. A zoom into the region around the tallest peaks is shown in the inset panel. }
   \label{fig:ls}
\end{figure}

To estimate $\psi$, we need to know the inclination of the stellar rotation axis with respect to the line of sight ($I_*$), which can be calculated if we know the rotation period of the star. The systematics-corrected light curve of \epic shows a clear quasi-periodic variability with a peak-to-peak amplitude of $\sim 600$ ppm. While we cannot be absolutely sure that this signal is induced by spot modulations rather than pulsations, since \epic resides in the typical \teff and \logg range for $\gamma$ Dor variables \citep[e.g.][]{bradley15}, the light curve strongly resembles those of spotted, rotating stars. We calculate a Lomb-Scargle periodogram \citep{lomb, scargle, zechmeister} of the \ktwo light curve with the transits masked out and low-frequency trends removed (see Fig.~\ref{fig:ls}). The periodogram shows three strong, closely-spaced peaks around $\sim 1.15$ days. But if we divide the light curve into segments of $\sim 10$ days in length and produce a periodogram for each, every periodogram shows a single, clear peak in the vicinity of 1.15 days. We adopt the mean of the peaks from the seven segments, resulting in a period of $1.152 \pm 0.023$ days. Again, the photometric variability may come from pulsations rather than spot modulations, but the rotation period of the star is related to the spectroscopically derived \vsini and stellar radius by $\sin I_\star = \vsini P_{rot}/2\pi R_\star$. A rotation period of 1.152 days would give us $\sin I_\star \simeq 1$. This coincidence gives us some confidence that the $1.152$-day peak is close to the true rotation period of the star.

To deduce the full three-dimensional spin-orbit alignment, we perform a Monte Carlo simulation with $2\times 10^6$ samples assuming independent Gaussian distributions for $\lambda$, $R_\star$, \vsini, the planet's orbital inclination and $P_{rot}$, and discarding any combination of parameters that leads to $\sin I_\star > 1$. The resulting distribution for $\psi$ implies that the true obliquity of the system is very close to zero, with $\psi = 3.7^{\circ +3.7^\circ}_{\ -1.8^\circ}$. For reference, the spin-orbit misalignment between the Earth and the Sun is about $7^\circ$, and that between Jupiter and the Sun is about $6^\circ$.

\begin{figure*}[htb] 
   \centering
   \includegraphics[width=\textwidth]{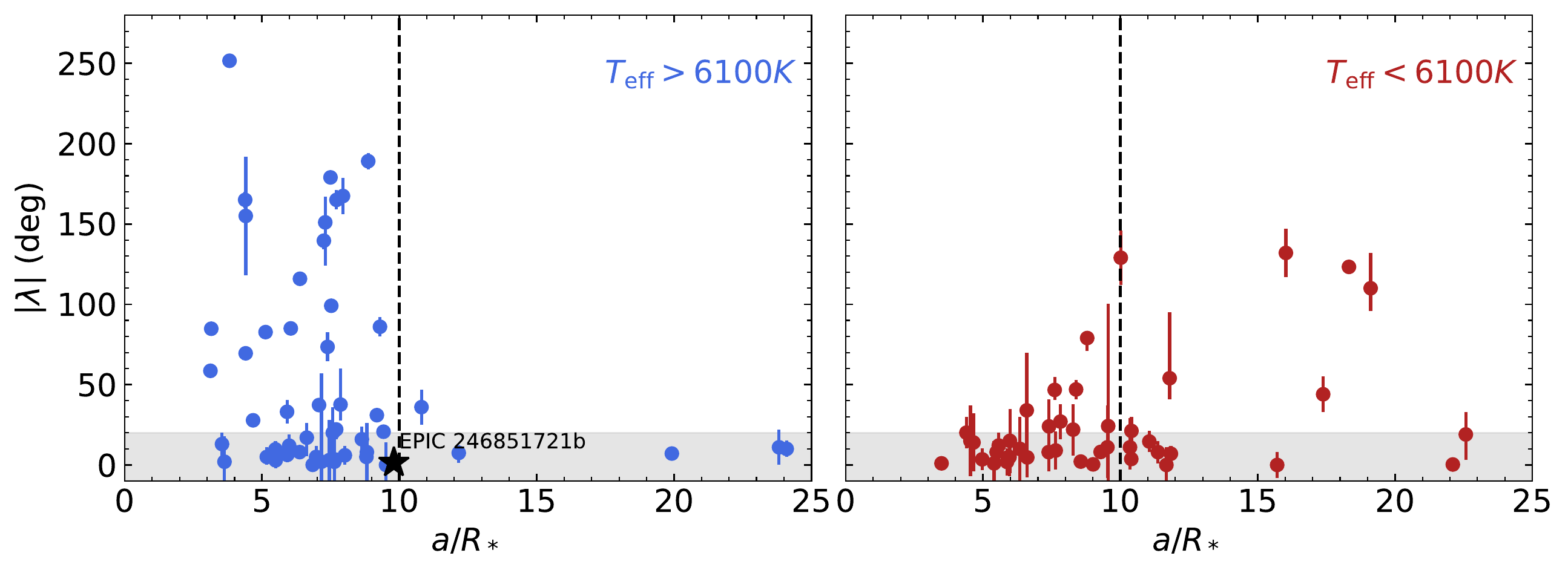} 
   \caption{Absolute values of the projected obliquities, $\lambda$, of Jovian planets as a function of scaled orbital distance, $a/R_*$, for systems with \teff $> 6100$ K (left) and \teff $< 6100$ K. The star shows the location of \planet. The grey shaded region corresponds to $|\lambda| < 20^\circ$ (aligned). Data retrieved from TEPCat\footnote{http://www.astro.keele.ac.uk/jkt/tepcat/obliquity.html} in June 2018.}
   \label{fig:hotcool}
\end{figure*}

Interestingly, six of the seven Jovian planets with \teff $>6100$ K, $P > 5$ days and known obliquities are well aligned ($|\lambda| < 20^\circ$), while only four out of ten such planets orbiting cooler stars are aligned. We can observe this apparent difference between the two populations more clearly in Fig.~\ref{fig:hotcool}, which shows the obliquity distributions of Jovian planets above and below the Kraft break as a function of $a/R_*$. The obliquity distributions at $a/R_* \gtrsim 10$ appear to contradict the empirical rule found by \citet{winn10} stating that stars with \teff $\gtrsim 6100$ K are more likely to host planets with high obliquities. \citet{albrecht12} and others reasoned that the original dichotomy found by \citet{winn10} is caused by differing amounts of mass in the stellar convective envelope, which acts to tidally dampen orbital obliquities. Cooler stars have a thicker convective envelope than hotter stars, resulting in stronger planet-star tidal interactions and shorter realignment timescales. 

But as we gather more planets with known obliquities at $a/R_* \gtrsim 10$, we observe that the fraction of misaligned planets around cooler stars starts to increase (right panel of Fig.~\ref{fig:hotcool}). This trend was first described by \citet{anderson15}, who reasoned that tides become ineffective with sufficient distance from the star, so the population at large $a/R_*$ may retain their initial spin-orbit angles instead of being tidally aligned. But \citet{anderson15} did not comment on the distribution around hot stars, presumably because there were too few planets with known obliquities at $a/R_* > 10$. Our sample of planets in that regime is limited, but they are mostly aligned, including \planet. We note that measured inner-disk-clearing timescales have been shown to be dependent on stellar mass \citep[e.g.][]{2009ApJ...700.1017K,2015A&A...576A..52R}, potentially due to the rate of photo-evaporation clearing the inner disk. A shorter disk clearing timescale for early-type stars may mean that the stopping distances of Jovian planets migrating in-disk are longer for early-type stellar hosts than for later-type hosts. These Jovian planets at larger semi-major axes may also be part of a group of giant planets that formed {\it in situ} \citep[e.g.][]{2016ApJ...817L..17B}, whose accretion timescales depend on the viscosity and lifetime of the disk. We must caution, however, that the current sample of early-type stars bearing well characterized Jovian planets at longer orbital periods is severely limited. A simple Student's T-test of the hot and cool star samples shows that the distinction between the two populations at distances of $a/R_*>10$ is not yet statistically significant. We expect this parameter space to be populated with planets from the \emph{TESS} mission, allowing a more statistically meaningful comparison to be made in the near future.

Is it possible that \planet initially had a higher obliquity and subsequently underwent tidal realignment? The \teff of \epic places it near the Kraft break \citep{kraft67}, the boundary between convective and radiative stars. The convective layer thins gradually with higher \teff, resulting in a smooth transition. So we consider tidal dissipation timescales for both convective and radiative stars \citep[Eqs. 2 and 3 from][]{albrecht12}:
\begin{eqnarray}
 &&   \frac{1}{\tau_\mathrm{CE}} = \frac{1}{10 \times 10^9 \ \mathrm{yr}} q^2 \Big( \frac{a/R_\star}{40}\Big)^{-6}, \\
 &&   \frac{1}{\tau_\mathrm{RA}}= \frac{1}{0.25 \times 5 \times 10^9 \ \mathrm{yr}} q^2 (1+q)^{5/6} \Big( \frac{a/R_\star}{6}\Big)^{-17/2}
\end{eqnarray}
where $\tau_\mathrm{CE}$ and $\tau_\mathrm{RA}$ are the tidal dissipation timescales for stars with convective and radiative envelopes respectively, and $q$ is the planet-to-star mass ratio ($M_P/M_\star$). For \planet, we estimate that $\tau_\mathrm{CE} \sim 10^{12}$ yrs and $\tau_\mathrm{RA} \sim 10^{16}$ yrs. Hence tidal dissipation is expected to be very weak whether the star is predominantly radiative or convective, though its rapid rotation suggests that it has a radiative envelope. Additionally, \citet{liwinn} also found that realignment through tidal effects would lead to a significant increase in the stellar rotation period, so a rapidly rotating star like \epic should not have undergone tidal realignment. The long tidal dissipation timescales, combined with the rapid rotation of the host star, suggest that \planet did not attain spin-orbit alignment through tidal interactions. It is therefore likely that the alignment we observe is primordial, which suggest that the planet formed {\it in situ} or underwent disk migration. This may be the case for all other aligned planets orbiting hot stars at $a/R_* \gtrsim 10$.
 
\subsection{Future Prospects}
The lack of planet-bearing early-type stars at longer orbital periods is a selection bias. Planets around the larger, early-type host stars at longer orbital periods are currently beyond the reach of ground-based surveys. With the successful launch of the Transiting Exoplanet Survey Satellite \citep[\tess,][]{ricker14}, we will soon begin to discover thousands of planets orbiting bright stars, including many Jovian planets with periods longer than 5 days, and sub-Jovian planets amenable to obliquity measurements. Both populations are relatively unexplored with respect to spin-orbit misalignment. A larger, more diverse sample may show us whether the primordial alignment in \epic is common among longer-period giant planets, and thereby reveal the dominant migration mechanism of close-in giant planets. Approximately 200 giant planets with periods longer than 5 days are expected to be discovered around stars brighter than $T_{\bf mag}=12$ in the TESS primary mission \citep{huang18}.  

\acknowledgments

We thank Elisabeth Newton and Jennifer Burt for helpful discussions on this manuscript. This research has made use of NASA's Astrophysics Data System, the Exoplanet Follow-up Observing Program (ExoFOP), and the NASA Exoplanet Archive, which is operated by the California Institute of Technology, under contract with the National Aeronautics and Space Administration under the Exoplanet Exploration Program. The National Geographic Society--Palomar Observatory Sky Atlas (POSS-I) was made by the California Institute of Technology with grants from the National Geographic Society. The Oschin Schmidt Telescope is operated by the California Institute of Technology and Palomar Observatory. 

This paper includes data collected by the \Kepler\ mission. Funding for the \Kepler\ mission is provided by the NASA Science Mission directorate. Some of the data presented in this paper were obtained from the Mikulski Archive for Space Telescopes (MAST). STScI is operated by the Association of Universities for Research in Astronomy, Inc., under NASA contract NAS5--26555. Support for MAST for non--HST data is provided by the NASA Office of Space Science via grant NNX13AC07G and by other grants and contracts.

This work has made use of data from the European Space Agency (ESA) mission
{\it Gaia} (\url{https://www.cosmos.esa.int/gaia}), processed by the {\it Gaia}
Data Processing and Analysis Consortium (DPAC, \url{https://www.cosmos.esa.int/web/gaia/dpac/consortium}). Funding for the DPAC
has been provided by national institutions, in particular the institutions
participating in the {\it Gaia} Multilateral Agreement. This publication makes use of data products from the Two Micron All Sky Survey, which is a joint project of the University of Massachusetts and the Infrared Processing and Analysis Center/California Institute of Technology, funded by the National Aeronautics and Space Administration and the National Science Foundation.

Data presented herein were obtained at the W. M. Keck Observatory from telescope time allocated to the National Aeronautics and Space Administration through the agency's scientific partnership with the California Institute of Technology and the University of California. The Observatory was made possible by the generous financial support of the W. M. Keck Foundation. The authors
wish to recognize and acknowledge the very significant cultural role
and reverence that the summit of Mauna Kea has always had within the
indigenous Hawaiian community.  We are most fortunate to have the
opportunity to conduct observations from this mountain. 

Work by G.Z. is provided by NASA through Hubble Fellowship grant HST-HF2-51402.001-A awarded by the Space Telescope Science Institute, which is operated by the Association of Universities for Research in Astronomy, Inc., for NASA, under contract NAS 5-26555. Work by C.H. is supported by the Juan Carlos Torres Fellowship. I.J.M.C. acknowledges support from NSF through grant AST-1824644. Work performed by J.E.R. is supported by the Harvard Future Faculty Leaders Postdoctoral fellowship.

Facilities: \facility{Kepler/K2, FLWO:1.5m (TRES), Keck-II (NIRC2), Magellan:Clay (MIKE)}

\bibliographystyle{apj_ads}

\begin{thebibliography}{}
\expandafter\ifx\csname natexlab\endcsname\relax\def\natexlab#1{#1}\fi

\bibitem[{{Adams} \& {Laughlin}(2006)}]{adamslaughlin}
{Adams}, F.~C., \& {Laughlin}, G. 2006,
  \href{http://dx.doi.org/10.1086/506145}{\apj},
  \href{http://adsabs.harvard.edu/abs/2006ApJ...649.1004A}{649},
  \href{http://adsabs.harvard.edu/abs/2006ApJ...649.1004A}{1004}

\bibitem[{{Albrecht} {et~al.}(2013){Albrecht}, {Winn}, {Marcy}, {Howard},
  {Isaacson}, \& {Johnson}}]{albrecht13}
{Albrecht}, S., {Winn}, J.~N., {Marcy}, G.~W., {et~al.} 2013,
  \href{http://dx.doi.org/10.1088/0004-637X/771/1/11}{\apj},
  \href{http://adsabs.harvard.edu/abs/2013ApJ...771...11A}{771},
  \href{http://adsabs.harvard.edu/abs/2013ApJ...771...11A}{11}

\bibitem[{{Albrecht} {et~al.}(2012){Albrecht}, {Winn}, {Johnson}, {Howard},
  {Marcy}, {Butler}, {Arriagada}, {Crane}, {Shectman}, {Thompson}, {Hirano},
  {Bakos}, \& {Hartman}}]{albrecht12}
{Albrecht}, S., {Winn}, J.~N., {Johnson}, J.~A., {et~al.} 2012,
  \href{http://dx.doi.org/10.1088/0004-637X/757/1/18}{\apj},
  \href{http://adsabs.harvard.edu/abs/2012ApJ...757...18A}{757},
  \href{http://adsabs.harvard.edu/abs/2012ApJ...757...18A}{18}

\bibitem[{{Alibert} {et~al.}(2005){Alibert}, {Mordasini}, {Benz}, \&
  {Winisdoerffer}}]{alibert05}
{Alibert}, Y., {Mordasini}, C., {Benz}, W., \& {Winisdoerffer}, C. 2005,
  \href{http://dx.doi.org/10.1051/0004-6361:20042032}{\aap},
  \href{http://adsabs.harvard.edu/abs/2005A\%26A...434..343A}{434},
  \href{http://adsabs.harvard.edu/abs/2005A\%26A...434..343A}{343}

\bibitem[{{Anderson} {et~al.}(2015){Anderson}, {Triaud}, {Turner}, {Brown},
  {Clark}, {Smalley}, {Collier Cameron}, {Doyle}, {Gillon}, {Hellier}, {Lovis},
  {Maxted}, {Pollacco}, {Queloz}, \& {Smith}}]{anderson15}
{Anderson}, D.~R., {Triaud}, A.~H.~M.~J., {Turner}, O.~D., {et~al.} 2015,
  \href{http://dx.doi.org/10.1088/2041-8205/800/1/L9}{\apjl},
  \href{http://adsabs.harvard.edu/abs/2015ApJ...800L...9A}{800},
  \href{http://adsabs.harvard.edu/abs/2015ApJ...800L...9A}{L9}

\bibitem[{{Batygin}(2012)}]{batygin12}
{Batygin}, K. 2012, \href{http://dx.doi.org/10.1038/nature11560}{\nat},
  \href{http://adsabs.harvard.edu/abs/2012Natur.491..418B}{491},
  \href{http://adsabs.harvard.edu/abs/2012Natur.491..418B}{418}

\bibitem[{{Batygin} \& {Adams}(2013)}]{Batygin:2013}
{Batygin}, K., \& {Adams}, F.~C. 2013,
  \href{http://dx.doi.org/10.1088/0004-637X/778/2/169}{\apj},
  \href{http://adsabs.harvard.edu/abs/2013ApJ...778..169B}{778},
  \href{http://adsabs.harvard.edu/abs/2013ApJ...778..169B}{169}

\bibitem[{{Batygin} {et~al.}(2016){Batygin}, {Bodenheimer}, \&
  {Laughlin}}]{batygin16}
{Batygin}, K., {Bodenheimer}, P.~H., \& {Laughlin}, G.~P. 2016,
  \href{http://dx.doi.org/10.3847/0004-637X/829/2/114}{\apj},
  \href{http://adsabs.harvard.edu/abs/2016ApJ...829..114B}{829},
  \href{http://adsabs.harvard.edu/abs/2016ApJ...829..114B}{114}

\bibitem[{{Bensby} {et~al.}(2003){Bensby}, {Feltzing}, \&
  {Lundstr{\"o}m}}]{Bensby:2003}
{Bensby}, T., {Feltzing}, S., \& {Lundstr{\"o}m}, I. 2003,
  \href{http://dx.doi.org/10.1051/0004-6361:20031213}{\aap},
  \href{http://adsabs.harvard.edu/abs/2003A\%26A...410..527B}{410},
  \href{http://adsabs.harvard.edu/abs/2003A\%26A...410..527B}{527}

\bibitem[{{Bernstein} {et~al.}(2003){Bernstein}, {Shectman}, {Gunnels},
  {Mochnacki}, \& {Athey}}]{Bernstein:2003}
{Bernstein}, R., {Shectman}, S.~A., {Gunnels}, S.~M., {Mochnacki}, S., \&
  {Athey}, A.~E. 2003, in \procspie, Vol. 4841, Instrument Design and
  Performance for Optical/Infrared Ground-based Telescopes, ed. M.~{Iye} \&
  A.~F.~M. {Moorwood}, 1694--1704, doi:10.1117/12.461502

\bibitem[{{Bieryla} {et~al.}(2015){Bieryla}, {Collins}, {Beatty}, {Eastman},
  {Siverd}, {Pepper}, {Gaudi}, {Stassun}, {Ca{\~n}as}, {Latham}, {Buchhave},
  {Sanchis-Ojeda}, {Winn}, {Jensen}, {Kielkopf}, {McLeod}, {Gregorio},
  {Col{\'o}n}, {Street}, {Ross}, {Penny}, {Mellon}, {Oberst}, {Fulton}, {Wang},
  {Berlind}, {Calkins}, {Esquerdo}, {DePoy}, {Gould}, {Marshall}, {Pogge},
  {Trueblood}, \& {Trueblood}}]{Bieryla:2015}
{Bieryla}, A., {Collins}, K., {Beatty}, T.~G., {et~al.} 2015,
  \href{http://dx.doi.org/10.1088/0004-6256/150/1/12}{\aj},
  \href{http://adsabs.harvard.edu/abs/2015AJ....150...12B}{150},
  \href{http://adsabs.harvard.edu/abs/2015AJ....150...12B}{12}

\bibitem[{{Bodenheimer} {et~al.}(2000){Bodenheimer}, {Hubickyj}, \&
  {Lissauer}}]{bodenheimer00}
{Bodenheimer}, P., {Hubickyj}, O., \& {Lissauer}, J.~J. 2000,
  \href{http://dx.doi.org/10.1006/icar.1999.6246}{Icarus},
  \href{http://adsabs.harvard.edu/abs/2000Icar..143....2B}{143},
  \href{http://adsabs.harvard.edu/abs/2000Icar..143....2B}{2}

\bibitem[{{Boley} {et~al.}(2016{\natexlab{a}}){Boley}, {Granados Contreras}, \&
  {Gladman}}]{Boley:2016}
{Boley}, A.~C., {Granados Contreras}, A.~P., \& {Gladman}, B.
  2016{\natexlab{a}},
  \href{http://dx.doi.org/10.3847/2041-8205/817/2/L17}{\apjl},
  \href{http://adsabs.harvard.edu/abs/2016ApJ...817L..17B}{817},
  \href{http://adsabs.harvard.edu/abs/2016ApJ...817L..17B}{L17}

\bibitem[{{Boley} {et~al.}(2016{\natexlab{b}}){Boley}, {Granados Contreras}, \&
  {Gladman}}]{2016ApJ...817L..17B}
---. 2016{\natexlab{b}},
  \href{http://dx.doi.org/10.3847/2041-8205/817/2/L17}{\apjl},
  \href{http://adsabs.harvard.edu/abs/2016ApJ...817L..17B}{817},
  \href{http://adsabs.harvard.edu/abs/2016ApJ...817L..17B}{L17}

\bibitem[{{Bourrier} {et~al.}(2015){Bourrier}, {Lecavelier des Etangs},
  {H{\'e}brard}, {Santerne}, {Deleuil}, {Almenara}, {Barros}, {Boisse},
  {Bonomo}, {Bruno}, {Courcol}, {Diaz}, {Montagnier}, \& {Moutou}}]{bourrier15}
{Bourrier}, V., {Lecavelier des Etangs}, A., {H{\'e}brard}, G., {et~al.} 2015,
  \href{http://dx.doi.org/10.1051/0004-6361/201525750}{\aap},
  \href{http://adsabs.harvard.edu/abs/2015A\%26A...579A..55B}{579},
  \href{http://adsabs.harvard.edu/abs/2015A\%26A...579A..55B}{A55}

\bibitem[{{Bovy}(2017)}]{bovy2017}
{Bovy}, J. 2017, \href{http://dx.doi.org/10.1093/mnras/stx1277}{\mnras},
  \href{http://adsabs.harvard.edu/abs/2017MNRAS.470.1360B}{470},
  \href{http://adsabs.harvard.edu/abs/2017MNRAS.470.1360B}{1360}

\bibitem[{{Bradley} {et~al.}(2015){Bradley}, {Guzik}, {Miles}, {Uytterhoeven},
  {Jackiewicz}, \& {Kinemuchi}}]{bradley15}
{Bradley}, P.~A., {Guzik}, J.~A., {Miles}, L.~F., {et~al.} 2015,
  \href{http://dx.doi.org/10.1088/0004-6256/149/2/68}{\aj},
  \href{http://adsabs.harvard.edu/abs/2015AJ....149...68B}{149},
  \href{http://adsabs.harvard.edu/abs/2015AJ....149...68B}{68}

\bibitem[{{Brown} {et~al.}(2012){Brown}, {Collier Cameron}, {D{\'{\i}}az},
  {Doyle}, {Gillon}, {Lendl}, {Smalley}, {Triaud}, {Anderson}, {Enoch},
  {Hellier}, {Maxted}, {Miller}, {Pollacco}, {Queloz}, {Boisse}, \&
  {H{\'e}brard}}]{brown12}
{Brown}, D.~J.~A., {Collier Cameron}, A., {D{\'{\i}}az}, R.~F., {et~al.} 2012,
  \href{http://dx.doi.org/10.1088/0004-637X/760/2/139}{\apj},
  \href{http://adsabs.harvard.edu/abs/2012ApJ...760..139B}{760},
  \href{http://adsabs.harvard.edu/abs/2012ApJ...760..139B}{139}

\bibitem[{{Buchhave} {et~al.}(2010){Buchhave}, {Bakos}, {Hartman}, {Torres},
  {Kov{\'a}cs}, {Latham}, {Noyes}, {Esquerdo}, {Everett}, {Howard}, {Marcy},
  {Fischer}, {Johnson}, {Andersen}, {F{\H u}r{\'e}sz}, {Perumpilly},
  {Sasselov}, {Stefanik}, {B{\'e}ky}, {L{\'a}z{\'a}r}, {Papp}, \&
  {S{\'a}ri}}]{buchhave2010}
{Buchhave}, L.~A., {Bakos}, G.~{\'A}., {Hartman}, J.~D., {et~al.} 2010,
  \href{http://dx.doi.org/10.1088/0004-637X/720/2/1118}{\apj},
  \href{http://adsabs.harvard.edu/abs/2010ApJ...720.1118B}{720},
  \href{http://adsabs.harvard.edu/abs/2010ApJ...720.1118B}{1118}

\bibitem[{{Buchhave} {et~al.}(2012){Buchhave}, {Latham}, {Johansen},
  {Bizzarro}, {Torres}, {Rowe}, {Batalha}, {Borucki}, {Brugamyer}, {Caldwell},
  {Bryson}, {Ciardi}, {Cochran}, {Endl}, {Esquerdo}, {Ford}, {Geary},
  {Gilliland}, {Hansen}, {Isaacson}, {Laird}, {Lucas}, {Marcy}, {Morse},
  {Robertson}, {Shporer}, {Stefanik}, {Still}, \& {Quinn}}]{buchhave12}
{Buchhave}, L.~A., {Latham}, D.~W., {Johansen}, A., {et~al.} 2012,
  \href{http://dx.doi.org/10.1038/nature11121}{\nat},
  \href{http://adsabs.harvard.edu/abs/2012Natur.486..375B}{486},
  \href{http://adsabs.harvard.edu/abs/2012Natur.486..375B}{375}

\bibitem[{{Castelli} \& {Kurucz}(2004)}]{CastelliKurucz:2004}
{Castelli}, F., \& {Kurucz}, R.~L. 2004, ArXiv Astrophysics e-prints,
  \href{http://adsabs.harvard.edu/abs/2004astro.ph..5087C}{arXiv:astro-ph/0405087}

\bibitem[{{Ciardi} {et~al.}(2015){Ciardi}, {Beichman}, {Horch}, \&
  {Howell}}]{ciardi2015}
{Ciardi}, D.~R., {Beichman}, C.~A., {Horch}, E.~P., \& {Howell}, S.~B. 2015,
  \href{http://dx.doi.org/10.1088/0004-637X/805/1/16}{\apj},
  \href{http://adsabs.harvard.edu/abs/2015ApJ...805...16C}{805},
  \href{http://adsabs.harvard.edu/abs/2015ApJ...805...16C}{16}

\bibitem[{{Ciardi} {et~al.}(2018){Ciardi}, {Crossfield}, {Feinstein},
  {Schlieder}, {Petigura}, {David}, {Bristow}, {Patel}, {Arnold}, {Benneke},
  {Christiansen}, {Dressing}, {Fulton}, {Howard}, {Isaacson}, {Sinukoff}, \&
  {Thackeray}}]{ciardi2018}
{Ciardi}, D.~R., {Crossfield}, I.~J.~M., {Feinstein}, A.~D., {et~al.} 2018,
  \href{http://dx.doi.org/10.3847/1538-3881/aa9921}{\aj},
  \href{http://adsabs.harvard.edu/abs/2018AJ....155...10C}{155},
  \href{http://adsabs.harvard.edu/abs/2018AJ....155...10C}{10}

\bibitem[{{Collier Cameron} {et~al.}(2010){Collier Cameron}, {Guenther},
  {Smalley}, {McDonald}, {Hebb}, {Andersen}, {Augusteijn}, {Barros}, {Brown},
  {Cochran}, {Endl}, {Fossey}, {Hartmann}, {Maxted}, {Pollacco}, {Skillen},
  {Telting}, {Waldmann}, \& {West}}]{Cameron:2010}
{Collier Cameron}, A., {Guenther}, E., {Smalley}, B., {et~al.} 2010,
  \href{http://dx.doi.org/10.1111/j.1365-2966.2010.16922.x}{\mnras},
  \href{http://adsabs.harvard.edu/abs/2010MNRAS.407..507C}{407},
  \href{http://adsabs.harvard.edu/abs/2010MNRAS.407..507C}{507}

\bibitem[{{Collins} {et~al.}(2017){Collins}, {Kielkopf}, {Stassun}, \&
  {Hessman}}]{collins2017}
{Collins}, K.~A., {Kielkopf}, J.~F., {Stassun}, K.~G., \& {Hessman}, F.~V.
  2017, \href{http://dx.doi.org/10.3847/1538-3881/153/2/77}{\aj},
  \href{http://adsabs.harvard.edu/abs/2017AJ....153...77C}{153},
  \href{http://adsabs.harvard.edu/abs/2017AJ....153...77C}{77}

\bibitem[{{Cutri} {et~al.}(2014){Cutri}, {Wright}, {Conrow}, {Fowler},
  {Eisenhardt}, {Grillmair}, {Kirkpatrick}, {Masci}, {McCallon}, {Wheelock},
  {Fajardo-Acosta}, {Yan}, {Benford}, {Harbut}, {Jarrett}, {Lake}, {Leisawitz},
  {Ressler}, {Stanford}, {Tsai}, {Liu}, {Helou}, {Mainzer}, {Gettings},
  {Gonzalez}, {Hoffman}, {Marsh}, {Padgett}, {Skrutskie}, {Beck}, {Papin}, \&
  {Wittman}}]{allwise}
{Cutri}, R.~M., {Wright}, E.~L., {Conrow}, T., {et~al.} 2014, VizieR Online
  Data Catalog, \href{http://cdsads.u-strasbg.fr/abs/2014yCat.2328....0C}{2328}

\bibitem[{{Damasso} {et~al.}(2015){Damasso}, {Esposito}, {Nascimbeni},
  {Desidera}, {Bonomo}, {Bieryla}, {Malavolta}, {Biazzo}, {Sozzetti}, {Covino},
  {Latham}, {Gandolfi}, {Rainer}, {Petrovich}, {Collins}, {Boccato}, {Claudi},
  {Cosentino}, {Gratton}, {Lanza}, {Maggio}, {Micela}, {Molinari}, {Pagano},
  {Piotto}, {Poretti}, {Smareglia}, {Di Fabrizio}, {Giacobbe}, {Gomez-Jimenez},
  {Murabito}, {Molinaro}, {Affer}, {Barbieri}, {Bedin}, {Benatti}, {Borsa},
  {Maldonado}, {Mancini}, {Scandariato}, {Southworth}, \& {Zanmar
  Sanchez}}]{damasso15}
{Damasso}, M., {Esposito}, M., {Nascimbeni}, V., {et~al.} 2015,
  \href{http://dx.doi.org/10.1051/0004-6361/201526995}{\aap},
  \href{http://adsabs.harvard.edu/abs/2015A\%26A...581L...6D}{581},
  \href{http://adsabs.harvard.edu/abs/2015A\%26A...581L...6D}{L6}

\bibitem[{{Damiani} \& {Lanza}(2015)}]{damianilanza}
{Damiani}, C., \& {Lanza}, A.~F. 2015,
  \href{http://dx.doi.org/10.1051/0004-6361/201424318}{\aap},
  \href{http://adsabs.harvard.edu/abs/2015A\%26A...574A..39D}{574},
  \href{http://adsabs.harvard.edu/abs/2015A\%26A...574A..39D}{A39}

\bibitem[{{Dawson} \& {Murray-Clay}(2013)}]{dawson13}
{Dawson}, R.~I., \& {Murray-Clay}, R.~A. 2013,
  \href{http://dx.doi.org/10.1088/2041-8205/767/2/L24}{\apjl},
  \href{http://adsabs.harvard.edu/abs/2013ApJ...767L..24D}{767},
  \href{http://adsabs.harvard.edu/abs/2013ApJ...767L..24D}{L24}

\bibitem[{{Donati} {et~al.}(1997){Donati}, {Semel}, {Carter}, {Rees}, \&
  {Collier Cameron}}]{Donati:1997}
{Donati}, J.-F., {Semel}, M., {Carter}, B.~D., {Rees}, D.~E., \& {Collier
  Cameron}, A. 1997, \href{http://dx.doi.org/10.1093/mnras/291.4.658}{\mnras},
  \href{http://adsabs.harvard.edu/abs/1997MNRAS.291..658D}{291},
  \href{http://adsabs.harvard.edu/abs/1997MNRAS.291..658D}{658}

\bibitem[{{Eastman}(2017)}]{Eastman:2017}
{Eastman}, J. 2017, {EXOFASTv2: Generalized publication-quality exoplanet
  modeling code}, Astrophysics Source Code Library,
  ascl:\href{http://adsabs.harvard.edu/abs/2017ascl.soft10003E}{1710.003}

\bibitem[{{Fabrycky} \& {Tremaine}(2007)}]{fabryckytremaine}
{Fabrycky}, D., \& {Tremaine}, S. 2007,
  \href{http://dx.doi.org/10.1086/521702}{\apj},
  \href{http://adsabs.harvard.edu/abs/2007ApJ...669.1298F}{669},
  \href{http://adsabs.harvard.edu/abs/2007ApJ...669.1298F}{1298}

\bibitem[{{Fabrycky}(2010)}]{fabrycky10}
{Fabrycky}, D.~C. 2010, ArXiv e-prints,
  arXiv:\href{http://adsabs.harvard.edu/abs/2010arXiv1006.3834F}{1006.3834}

\bibitem[{Fur{\'e}sz(2008)}]{Furesz:2008}
Fur{\'e}sz, G. 2008, PhD thesis, University of Szeged

\bibitem[{{Furlan} {et~al.}(2017){Furlan}, {Ciardi}, {Everett}, {Saylors},
  {Teske}, {Horch}, {Howell}, {van Belle}, {Hirsch}, {Gautier}, {Adams},
  {Barrado}, {Cartier}, {Dressing}, {Dupree}, {Gilliland}, {Lillo-Box},
  {Lucas}, \& {Wang}}]{furlan2017}
{Furlan}, E., {Ciardi}, D.~R., {Everett}, M.~E., {et~al.} 2017,
  \href{http://dx.doi.org/10.3847/1538-3881/153/2/71}{\aj},
  \href{http://adsabs.harvard.edu/abs/2017AJ....153...71F}{153},
  \href{http://adsabs.harvard.edu/abs/2017AJ....153...71F}{71}

\bibitem[{{Gaia Collaboration} {et~al.}(2018){Gaia Collaboration}, {Brown},
  {Vallenari}, {Prusti}, {de Bruijne}, {Babusiaux}, \& {Bailer-Jones}}]{gaia18}
{Gaia Collaboration}, {Brown}, A.~G.~A., {Vallenari}, A., {et~al.} 2018, ArXiv
  e-prints,
  arXiv:\href{http://adsabs.harvard.edu/abs/2018arXiv180409365G}{1804.09365}

\bibitem[{{Gaia Collaboration} {et~al.}(2016){Gaia Collaboration}, {Brown},
  {Vallenari}, {Prusti}, {de Bruijne}, {Mignard}, {Drimmel}, {Babusiaux},
  {Bailer-Jones}, {Bastian}, \& et~al.}]{gaia16}
---. 2016, \href{http://dx.doi.org/10.1051/0004-6361/201629512}{\aap},
  \href{http://adsabs.harvard.edu/abs/2016A\%26A...595A...2G}{595},
  \href{http://adsabs.harvard.edu/abs/2016A\%26A...595A...2G}{A2}

\bibitem[{{Gray}(2005)}]{Graybook:2005}
{Gray}, D.~F. 2005, {The Observation and Analysis of Stellar Photospheres}

\bibitem[{{Gray} \& {Corbally}(1994)}]{Gray:1994}
{Gray}, R.~O., \& {Corbally}, C.~J. 1994,
  \href{http://dx.doi.org/10.1086/116893}{\aj},
  \href{http://adsabs.harvard.edu/abs/1994AJ....107..742G}{107},
  \href{http://adsabs.harvard.edu/abs/1994AJ....107..742G}{742}

\bibitem[{{Hartman} \& {Bakos}(2016)}]{vartools}
{Hartman}, J.~D., \& {Bakos}, G.~{\'A}. 2016,
  \href{http://dx.doi.org/10.1016/j.ascom.2016.05.006}{Astronomy and
  Computing}, \href{http://adsabs.harvard.edu/abs/2016A\%26C....17....1H}{17},
  \href{http://adsabs.harvard.edu/abs/2016A\%26C....17....1H}{1}

\bibitem[{{H{\o}g} {et~al.}(2000){H{\o}g}, {Fabricius}, {Makarov}, {Urban},
  {Corbin}, {Wycoff}, {Bastian}, {Schwekendiek}, \& {Wicenec}}]{tycho}
{H{\o}g}, E., {Fabricius}, C., {Makarov}, V.~V., {et~al.} 2000, \aap,
  \href{http://adsabs.harvard.edu/abs/2000A\%26A...355L..27H}{355},
  \href{http://adsabs.harvard.edu/abs/2000A\%26A...355L..27H}{L27}

\bibitem[{{Howell} {et~al.}(2012){Howell}, {Rowe}, {Bryson}, {Quinn}, {Marcy},
  {Isaacson}, {Ciardi}, {Chaplin}, {Metcalfe}, {Monteiro}, {Appourchaux},
  {Basu}, {Creevey}, {Gilliland}, {Quirion}, {Stello}, {Kjeldsen},
  {Christensen-Dalsgaard}, {Elsworth}, {Garc{\'{\i}}a}, {Houdek}, {Karoff},
  {Molenda-{\.Z}akowicz}, {Thompson}, {Verner}, {Torres}, {Fressin}, {Crepp},
  {Adams}, {Dupree}, {Sasselov}, {Dressing}, {Borucki}, {Koch}, {Lissauer},
  {Latham}, {Buchhave}, {Gautier}, {Everett}, {Horch}, {Batalha}, {Dunham},
  {Szkody}, {Silva}, {Mighell}, {Holberg}, {Ballot}, {Bedding}, {Bruntt},
  {Campante}, {Handberg}, {Hekker}, {Huber}, {Mathur}, {Mosser}, {R{\'e}gulo},
  {White}, {Christiansen}, {Middour}, {Haas}, {Hall}, {Jenkins}, {McCaulif},
  {Fanelli}, {Kulesa}, {McCarthy}, \& {Henze}}]{howell12}
{Howell}, S.~B., {Rowe}, J.~F., {Bryson}, S.~T., {et~al.} 2012,
  \href{http://dx.doi.org/10.1088/0004-637X/746/2/123}{\apj},
  \href{http://adsabs.harvard.edu/abs/2012ApJ...746..123H}{746},
  \href{http://adsabs.harvard.edu/abs/2012ApJ...746..123H}{123}

\bibitem[{{Huang} {et~al.}(2018){Huang}, {Shporer}, {Dragomir}, {Fausnaugh},
  {Levine}, {Morgan}, {Nguyen}, {Ricker}, {Wall}, {Woods}, \&
  {Vanderspek}}]{huang18}
{Huang}, C.~X., {Shporer}, A., {Dragomir}, D., {et~al.} 2018, ArXiv e-prints,
  arXiv:\href{http://adsabs.harvard.edu/abs/2018arXiv180711129H}{1807.11129}

\bibitem[{{Huang} {et~al.}(2013){Huang}, {Bakos}, \& {Hartman}}]{huang13}
{Huang}, X., {Bakos}, G.~{\'A}., \& {Hartman}, J.~D. 2013,
  \href{http://dx.doi.org/10.1093/mnras/sts463}{\mnras},
  \href{http://adsabs.harvard.edu/abs/2013MNRAS.429.2001H}{429},
  \href{http://adsabs.harvard.edu/abs/2013MNRAS.429.2001H}{2001}

\bibitem[{{Huang} {et~al.}(2015){Huang}, {Penev}, {Hartman}, {Bakos}, {Bhatti},
  {Domsa}, \& {de Val-Borro}}]{huang15}
{Huang}, X., {Penev}, K., {Hartman}, J.~D., {et~al.} 2015,
  \href{http://dx.doi.org/10.1093/mnras/stv2257}{\mnras},
  \href{http://adsabs.harvard.edu/abs/2015MNRAS.454.4159H}{454},
  \href{http://adsabs.harvard.edu/abs/2015MNRAS.454.4159H}{4159}

\bibitem[{{Kelson}(2003)}]{2003PASP..115..688K}
{Kelson}, D.~D. 2003, \href{http://dx.doi.org/10.1086/375502}{\pasp},
  \href{http://adsabs.harvard.edu/abs/2003PASP..115..688K}{115},
  \href{http://adsabs.harvard.edu/abs/2003PASP..115..688K}{688}

\bibitem[{{Kelson} {et~al.}(2000){Kelson}, {Illingworth}, {van Dokkum}, \&
  {Franx}}]{2000ApJ...531..159K}
{Kelson}, D.~D., {Illingworth}, G.~D., {van Dokkum}, P.~G., \& {Franx}, M.
  2000, \href{http://dx.doi.org/10.1086/308445}{\apj},
  \href{http://adsabs.harvard.edu/abs/2000ApJ...531..159K}{531},
  \href{http://adsabs.harvard.edu/abs/2000ApJ...531..159K}{159}

\bibitem[{{Kim} {et~al.}(2009){Kim}, {Watson}, {Manoj}, {Furlan}, {Najita},
  {Forrest}, {Sargent}, {Espaillat}, {Calvet}, {Luhman}, {McClure}, {Green}, \&
  {Harrold}}]{2009ApJ...700.1017K}
{Kim}, K.~H., {Watson}, D.~M., {Manoj}, P., {et~al.} 2009,
  \href{http://dx.doi.org/10.1088/0004-637X/700/2/1017}{\apj},
  \href{http://adsabs.harvard.edu/abs/2009ApJ...700.1017K}{700},
  \href{http://adsabs.harvard.edu/abs/2009ApJ...700.1017K}{1017}

\bibitem[{{Kiseleva} {et~al.}(1998){Kiseleva}, {Eggleton}, \&
  {Mikkola}}]{kiseleva98}
{Kiseleva}, L.~G., {Eggleton}, P.~P., \& {Mikkola}, S. 1998,
  \href{http://dx.doi.org/10.1046/j.1365-8711.1998.01903.x}{\mnras},
  \href{http://adsabs.harvard.edu/abs/1998MNRAS.300..292K}{300},
  \href{http://adsabs.harvard.edu/abs/1998MNRAS.300..292K}{292}

\bibitem[{{Kov{\'a}cs} {et~al.}(2002){Kov{\'a}cs}, {Zucker}, \& {Mazeh}}]{bls}
{Kov{\'a}cs}, G., {Zucker}, S., \& {Mazeh}, T. 2002,
  \href{http://dx.doi.org/10.1051/0004-6361:20020802}{\aap},
  \href{http://adsabs.harvard.edu/abs/2002A\%26A...391..369K}{391},
  \href{http://adsabs.harvard.edu/abs/2002A\%26A...391..369K}{369}

\bibitem[{{Kraft}(1967)}]{kraft67}
{Kraft}, R.~P. 1967, \href{http://dx.doi.org/10.1086/149359}{\apj},
  \href{http://adsabs.harvard.edu/abs/1967ApJ...150..551K}{150},
  \href{http://adsabs.harvard.edu/abs/1967ApJ...150..551K}{551}

\bibitem[{{Lai}(2012)}]{lai12}
{Lai}, D. 2012,
  \href{http://dx.doi.org/10.1111/j.1365-2966.2012.20893.x}{\mnras},
  \href{http://adsabs.harvard.edu/abs/2012MNRAS.423..486L}{423},
  \href{http://adsabs.harvard.edu/abs/2012MNRAS.423..486L}{486}

\bibitem[{{Lai}(2014)}]{lai14}
---. 2014, \href{http://dx.doi.org/10.1093/mnras/stu485}{\mnras},
  \href{http://adsabs.harvard.edu/abs/2014MNRAS.440.3532L}{440},
  \href{http://adsabs.harvard.edu/abs/2014MNRAS.440.3532L}{3532}

\bibitem[{{Li} \& {Winn}(2016)}]{liwinn}
{Li}, G., \& {Winn}, J.~N. 2016,
  \href{http://dx.doi.org/10.3847/0004-637X/818/1/5}{\apj},
  \href{http://adsabs.harvard.edu/abs/2016ApJ...818....5L}{818},
  \href{http://adsabs.harvard.edu/abs/2016ApJ...818....5L}{5}

\bibitem[{{Lin} {et~al.}(1996){Lin}, {Bodenheimer}, \& {Richardson}}]{lin96}
{Lin}, D.~N.~C., {Bodenheimer}, P., \& {Richardson}, D.~C. 1996,
  \href{http://dx.doi.org/10.1038/380606a0}{\nat},
  \href{http://adsabs.harvard.edu/abs/1996Natur.380..606L}{380},
  \href{http://adsabs.harvard.edu/abs/1996Natur.380..606L}{606}

\bibitem[{{Lin} \& {Ogilvie}(2017)}]{linogilvie}
{Lin}, Y., \& {Ogilvie}, G.~I. 2017,
  \href{http://dx.doi.org/10.1093/mnras/stx540}{\mnras},
  \href{http://adsabs.harvard.edu/abs/2017MNRAS.468.1387L}{468},
  \href{http://adsabs.harvard.edu/abs/2017MNRAS.468.1387L}{1387}

\bibitem[{{Lomb}(1976)}]{lomb}
{Lomb}, N.~R. 1976, \href{http://dx.doi.org/10.1007/BF00648343}{\apss},
  \href{http://adsabs.harvard.edu/abs/1976Ap\%26SS..39..447L}{39},
  \href{http://adsabs.harvard.edu/abs/1976Ap\%26SS..39..447L}{447}

\bibitem[{{Mazeh} {et~al.}(2015){Mazeh}, {Perets}, {McQuillan}, \&
  {Goldstein}}]{mazeh15}
{Mazeh}, T., {Perets}, H.~B., {McQuillan}, A., \& {Goldstein}, E.~S. 2015,
  \href{http://dx.doi.org/10.1088/0004-637X/801/1/3}{\apj},
  \href{http://adsabs.harvard.edu/abs/2015ApJ...801....3M}{801},
  \href{http://adsabs.harvard.edu/abs/2015ApJ...801....3M}{3}

\bibitem[{{McLaughlin}(1924)}]{McLaughlin:1924}
{McLaughlin}, D.~B. 1924, \href{http://dx.doi.org/10.1086/142826}{\apj},
  \href{http://adsabs.harvard.edu/abs/1924ApJ....60...22M}{60},
  \href{http://adsabs.harvard.edu/abs/1924ApJ....60...22M}{22}

\bibitem[{{Morton}(2015)}]{isochrones}
{Morton}, T.~D. 2015, {isochrones: Stellar model grid package}, Astrophysics
  Source Code Library,
  ascl:\href{http://adsabs.harvard.edu/abs/2015ascl.soft03010M}{1503.010}

\bibitem[{{Naoz} {et~al.}(2011){Naoz}, {Farr}, {Lithwick}, {Rasio}, \&
  {Teyssandier}}]{naoz11}
{Naoz}, S., {Farr}, W.~M., {Lithwick}, Y., {Rasio}, F.~A., \& {Teyssandier}, J.
  2011, \href{http://dx.doi.org/10.1038/nature10076}{\nat},
  \href{http://adsabs.harvard.edu/abs/2011Natur.473..187N}{473},
  \href{http://adsabs.harvard.edu/abs/2011Natur.473..187N}{187}

\bibitem[{{Ngo} {et~al.}(2016){Ngo}, {Knutson}, {Hinkley}, {Bryan}, {Crepp},
  {Batygin}, {Crossfield}, {Hansen}, {Howard}, {Johnson}, {Mawet}, {Morton},
  {Muirhead}, \& {Wang}}]{ngo16}
{Ngo}, H., {Knutson}, H.~A., {Hinkley}, S., {et~al.} 2016,
  \href{http://dx.doi.org/10.3847/0004-637X/827/1/8}{\apj},
  \href{http://adsabs.harvard.edu/abs/2016ApJ...827....8N}{827},
  \href{http://adsabs.harvard.edu/abs/2016ApJ...827....8N}{8}

\bibitem[{{Pecaut} \& {Mamajek}(2013)}]{pecautmamajek13}
{Pecaut}, M.~J., \& {Mamajek}, E.~E. 2013,
  \href{http://dx.doi.org/10.1088/0067-0049/208/1/9}{\apjs},
  \href{http://adsabs.harvard.edu/abs/2013ApJS..208....9P}{208},
  \href{http://adsabs.harvard.edu/abs/2013ApJS..208....9P}{9}

\bibitem[{{Rasio} \& {Ford}(1996)}]{rasioford}
{Rasio}, F.~A., \& {Ford}, E.~B. 1996,
  \href{http://dx.doi.org/10.1126/science.274.5289.954}{Science},
  \href{http://adsabs.harvard.edu/abs/1996Sci...274..954R}{274},
  \href{http://adsabs.harvard.edu/abs/1996Sci...274..954R}{954}

\bibitem[{{Ribas} {et~al.}(2015){Ribas}, {Bouy}, \&
  {Mer{\'{\i}}n}}]{2015A&A...576A..52R}
{Ribas}, {\'A}., {Bouy}, H., \& {Mer{\'{\i}}n}, B. 2015,
  \href{http://dx.doi.org/10.1051/0004-6361/201424846}{\aap},
  \href{http://adsabs.harvard.edu/abs/2015A%26A...576A..52R}{576},
  \href{http://adsabs.harvard.edu/abs/2015A%26A...576A..52R}{A52}

\bibitem[{{Ricker} {et~al.}(2014){Ricker}, {Winn}, {Vanderspek}, {Latham},
  {Bakos}, {Bean}, {Berta-Thompson}, {Brown}, {Buchhave}, {Butler}, {Butler},
  {Chaplin}, {Charbonneau}, {Christensen-Dalsgaard}, {Clampin}, {Deming},
  {Doty}, {De Lee}, {Dressing}, {Dunham}, {Endl}, {Fressin}, {Ge}, {Henning},
  {Holman}, {Howard}, {Ida}, {Jenkins}, {Jernigan}, {Johnson}, {Kaltenegger},
  {Kawai}, {Kjeldsen}, {Laughlin}, {Levine}, {Lin}, {Lissauer}, {MacQueen},
  {Marcy}, {McCullough}, {Morton}, {Narita}, {Paegert}, {Palle}, {Pepe},
  {Pepper}, {Quirrenbach}, {Rinehart}, {Sasselov}, {Sato}, {Seager},
  {Sozzetti}, {Stassun}, {Sullivan}, {Szentgyorgyi}, {Torres}, {Udry}, \&
  {Villasenor}}]{ricker14}
{Ricker}, G.~R., {Winn}, J.~N., {Vanderspek}, R., {et~al.} 2014, in \procspie,
  Vol. 9143, Space Telescopes and Instrumentation 2014: Optical, Infrared, and
  Millimeter Wave, 914320,
  arXiv:\href{http://adsabs.harvard.edu/abs/2014SPIE.9143E..20R}{1406.0151}

\bibitem[{{Rossiter}(1924)}]{Rossiter:1924}
{Rossiter}, R.~A. 1924, \href{http://dx.doi.org/10.1086/142825}{\apj},
  \href{http://adsabs.harvard.edu/abs/1924ApJ....60...15R}{60},
  \href{http://adsabs.harvard.edu/abs/1924ApJ....60...15R}{15}

\bibitem[{{Sanchis-Ojeda} {et~al.}(2011){Sanchis-Ojeda}, {Winn}, {Holman},
  {Carter}, {Osip}, \& {Fuentes}}]{sanchisojeda11}
{Sanchis-Ojeda}, R., {Winn}, J.~N., {Holman}, M.~J., {et~al.} 2011,
  \href{http://dx.doi.org/10.1088/0004-637X/733/2/127}{\apj},
  \href{http://adsabs.harvard.edu/abs/2011ApJ...733..127S}{733},
  \href{http://adsabs.harvard.edu/abs/2011ApJ...733..127S}{127}

\bibitem[{{Scargle}(1982)}]{scargle}
{Scargle}, J.~D. 1982, \href{http://dx.doi.org/10.1086/160554}{\apj},
  \href{http://adsabs.harvard.edu/abs/1982ApJ...263..835S}{263},
  \href{http://adsabs.harvard.edu/abs/1982ApJ...263..835S}{835}

\bibitem[{{Schlaufman}(2010)}]{schlaufman10}
{Schlaufman}, K.~C. 2010,
  \href{http://dx.doi.org/10.1088/0004-637X/719/1/602}{\apj},
  \href{http://adsabs.harvard.edu/abs/2010ApJ...719..602S}{719},
  \href{http://adsabs.harvard.edu/abs/2010ApJ...719..602S}{602}

\bibitem[{{Schlegel} {et~al.}(1998){Schlegel}, {Finkbeiner}, \&
  {Davis}}]{Schlegel1998}
{Schlegel}, D.~J., {Finkbeiner}, D.~P., \& {Davis}, M. 1998,
  \href{http://dx.doi.org/10.1086/305772}{\apj},
  \href{http://adsabs.harvard.edu/abs/1998ApJ...500..525S}{500},
  \href{http://adsabs.harvard.edu/abs/1998ApJ...500..525S}{525}

\bibitem[{{Skrutskie} {et~al.}(2006){Skrutskie}, {Cutri}, {Stiening},
  {Weinberg}, {Schneider}, {Carpenter}, {Beichman}, {Capps}, {Chester},
  {Elias}, {Huchra}, {Liebert}, {Lonsdale}, {Monet}, {Price}, {Seitzer},
  {Jarrett}, {Kirkpatrick}, {Gizis}, {Howard}, {Evans}, {Fowler}, {Fullmer},
  {Hurt}, {Light}, {Kopan}, {Marsh}, {McCallon}, {Tam}, {Van Dyk}, \&
  {Wheelock}}]{twomass}
{Skrutskie}, M.~F., {Cutri}, R.~M., {Stiening}, R., {et~al.} 2006,
  \href{http://dx.doi.org/10.1086/498708}{\aj},
  \href{http://adsabs.harvard.edu/abs/2006AJ....131.1163S}{131},
  \href{http://adsabs.harvard.edu/abs/2006AJ....131.1163S}{1163}

\bibitem[{{Stassun} \& {Torres}(2018)}]{stassun2018}
{Stassun}, K.~G., \& {Torres}, G. 2018, ArXiv e-prints,
  arXiv:\href{http://adsabs.harvard.edu/abs/2018arXiv180503526S}{1805.03526}

\bibitem[{{Szentgyorgyi} \& {Fur{\'e}sz}(2007)}]{tres}
{Szentgyorgyi}, A.~H., \& {Fur{\'e}sz}, G. 2007, in Revista Mexicana de
  Astronomia y Astrofisica, vol.~27, Vol.~28, Revista Mexicana de Astronomia y
  Astrofisica Conference Series, ed. S.~{Kurtz}, 129--133

\bibitem[{{Vanderburg} \& {Johnson}(2014)}]{vj14}
{Vanderburg}, A., \& {Johnson}, J.~A. 2014,
  \href{http://dx.doi.org/10.1086/678764}{\pasp},
  \href{http://adsabs.harvard.edu/abs/2014PASP..126..948V}{126},
  \href{http://adsabs.harvard.edu/abs/2014PASP..126..948V}{948}

\bibitem[{{Vanderburg} {et~al.}(2016){Vanderburg}, {Latham}, {Buchhave},
  {Bieryla}, {Berlind}, {Calkins}, {Esquerdo}, {Welsh}, \& {Johnson}}]{v16}
{Vanderburg}, A., {Latham}, D.~W., {Buchhave}, L.~A., {et~al.} 2016,
  \href{http://dx.doi.org/10.3847/0067-0049/222/1/14}{\apjs},
  \href{http://adsabs.harvard.edu/abs/2016ApJS..222...14V}{222},
  \href{http://adsabs.harvard.edu/abs/2016ApJS..222...14V}{14}

\bibitem[{{Villanueva} {et~al.}(2018){Villanueva}, {Gaudi}, {Pogge}, {Eastman},
  {Stassun}, {Trueblood}, \& {Trueblood}}]{demonext}
{Villanueva}, Jr., S., {Gaudi}, B.~S., {Pogge}, R.~W., {et~al.} 2018,
  \href{http://dx.doi.org/10.1088/1538-3873/aa9603}{\pasp},
  \href{http://adsabs.harvard.edu/abs/2018PASP..130a5001V}{130},
  \href{http://adsabs.harvard.edu/abs/2018PASP..130a5001V}{015001}

\bibitem[{{Winn} {et~al.}(2010){Winn}, {Fabrycky}, {Albrecht}, \&
  {Johnson}}]{winn10}
{Winn}, J.~N., {Fabrycky}, D., {Albrecht}, S., \& {Johnson}, J.~A. 2010,
  \href{http://dx.doi.org/10.1088/2041-8205/718/2/L145}{\apjl},
  \href{http://adsabs.harvard.edu/abs/2010ApJ...718L.145W}{718},
  \href{http://adsabs.harvard.edu/abs/2010ApJ...718L.145W}{L145}

\bibitem[{{Winn} {et~al.}(2007){Winn}, {Johnson}, {Peek}, {Marcy}, {Bakos},
  {Enya}, {Narita}, {Suto}, {Turner}, \& {Vogt}}]{winn07}
{Winn}, J.~N., {Johnson}, J.~A., {Peek}, K.~M.~G., {et~al.} 2007,
  \href{http://dx.doi.org/10.1086/521362}{\apjl},
  \href{http://adsabs.harvard.edu/abs/2007ApJ...665L.167W}{665},
  \href{http://adsabs.harvard.edu/abs/2007ApJ...665L.167W}{L167}

\bibitem[{{Wu} \& {Murray}(2003)}]{wumurray}
{Wu}, Y., \& {Murray}, N. 2003, \href{http://dx.doi.org/10.1086/374598}{\apj},
  \href{http://adsabs.harvard.edu/abs/2003ApJ...589..605W}{589},
  \href{http://adsabs.harvard.edu/abs/2003ApJ...589..605W}{605}

\bibitem[{{Yi} {et~al.}(2001){Yi}, {Demarque}, {Kim}, {Lee}, {Ree}, {Lejeune},
  \& {Barnes}}]{Yi:2001}
{Yi}, S., {Demarque}, P., {Kim}, Y.-C., {et~al.} 2001,
  \href{http://dx.doi.org/10.1086/321795}{\apjs},
  \href{http://adsabs.harvard.edu/abs/2001ApJS..136..417Y}{136},
  \href{http://adsabs.harvard.edu/abs/2001ApJS..136..417Y}{417}

\bibitem[{{Zechmeister} \& {K{\"u}rster}(2009)}]{zechmeister}
{Zechmeister}, M., \& {K{\"u}rster}, M. 2009,
  \href{http://dx.doi.org/10.1051/0004-6361:200811296}{\aap},
  \href{http://adsabs.harvard.edu/abs/2009A\%26A...496..577Z}{496},
  \href{http://adsabs.harvard.edu/abs/2009A\%26A...496..577Z}{577}

\bibitem[{{Zhou} {et~al.}(2016){Zhou}, {Latham}, {Bieryla}, {Beatty},
  {Buchhave}, {Esquerdo}, {Berlind}, \& {Calkins}}]{Zhou:2016a}
{Zhou}, G., {Latham}, D.~W., {Bieryla}, A., {et~al.} 2016,
  \href{http://dx.doi.org/10.1093/mnras/stw1107}{\mnras},
  \href{http://adsabs.harvard.edu/abs/2016MNRAS.460.3376Z}{460},
  \href{http://adsabs.harvard.edu/abs/2016MNRAS.460.3376Z}{3376}

\bibitem[{{Zhou} {et~al.}(2018){Zhou}, {Rodriguez}, {Vanderburg}, {Quinn},
  {Irwin}, {Huang}, {Latham}, {Bieryla}, {Esquerdo}, {Berlind}, \&
  {Calkins}}]{Zhou18}
{Zhou}, G., {Rodriguez}, J.~E., {Vanderburg}, A., {et~al.} 2018,
  \href{http://dx.doi.org/10.3847/1538-3881/aad085}{\aj},
  \href{http://adsabs.harvard.edu/abs/2018AJ....156...93Z}{156},
  \href{http://adsabs.harvard.edu/abs/2018AJ....156...93Z}{93}

\end{thebibliography}

\end{document}